\documentclass[twocolumn]{aastex631}

\usepackage{graphicx}
\usepackage[none]{hyphenat}
\usepackage{amsmath}
\usepackage{booktabs}
\usepackage{multirow}
\usepackage{txfonts}
\usepackage{enumitem}

\usepackage{algorithmic}
\usepackage{bm}
\usepackage{mathrsfs}
\usepackage{booktabs}  
\usepackage{float}
\usepackage{subfigure}
\DeclareRobustCommand{\VAN}[3]{#2}
\let\VANthebibliography\thebibliography
\def\thebibliography{\DeclareRobustCommand{\VAN}[3]{##3}\VANthebibliography}
\usepackage{color}
\shorttitle{Particle Acceleration and Diffusion}
\shortauthors{Gao \& Zhang}

\begin{document}
\title{The diffusion and scattering of accelerating particles in compressible MHD turbulence}

\author{Na-Na Gao}
\affiliation{Department of Physics, Xiangtan University, Xiangtan, Hunan 411105, China}

\author{Jian-Fu Zhang}
\affiliation{Department of Physics, Xiangtan University, Xiangtan, Hunan 411105, China}
\affiliation{Key Laboratory of Stars and Interstellar Medium, Xiangtan University, Xiangtan 411105, China}
\affiliation{Department of Astronomy and Space Science, Chungnam National University, Daejeon, Republic of Korea}
\email{jfzhang@xtu.edu.cn}

\begin{abstract}
We numerically study the diffusion and scattering of cosmic rays (CRs) together with their acceleration processes in the framework of the modern understanding of magnetohydrodynamic (MHD) turbulence. Based on the properties of compressible MHD turbulence obtained from observations and numerical experiments, we investigate the interaction of CRs with plasma modes. We find that (1) the gyroradius of particles exponentially increases with the acceleration timescale; (2) the momentum diffusion presents the power-law relationship with the gyroradius in the strong turbulence regime, and shows a plateau in the weak turbulence regime implying a stochastic acceleration process; (3) the spatial diffusion is dominated by the parallel diffusion in the sub-Alfv\'enic regime, while it is dominated by the perpendicular diffusion in the super-Alfv\'enic one; (4) as for the interaction of CRs with plasma modes, the particle acceleration is dominated by the fast mode in the high $\beta$ case, while in the low $\beta$ case, it is dominated by the fast and slow modes; (5) in the presence of acceleration, magnetosonic modes still play a critical role in diffusion and scattering processes of CRs, which is in good agreement with the earlier theoretical predictions. 
\end{abstract}


\section{Introduction}\label{sec:intro}
The propagation of cosmic rays (CRs), including CRs scattering, diffusion, and acceleration/re-acceleration processes, plays a crucial role in understanding high-energy phenomena from astrophysical sources. It was pointed out that magnetohydrodynamic (MHD) turbulence affects directly or indirectly the propagation of the CRs. We can safely say that MHD turbulence is an essential agent for pitch-angle scattering (\citealt{Yan2002}), spatial diffusion (\citealt{Casse2001,Yan2008,Xu2013,Lazarian2021,Maiti2022}), stochastic acceleration (\citealt{Fermi1949, Zhang2021}), and reconnection acceleration (\citealt{Lazarian1999}, hereafter LV99). Studying the particle acceleration mechanisms and understanding the diffusion of CRs in a general MHD turbulence can help us to comprehend the roles that CRs play in many key and complex astrophysical environments, such as solar physics (\citealt{Petrosian2008, Yan2008, Bian2012}), active galactic nucleus (AGN, \citealt{De2013, Mbarek2022}), gamma-ray bursts (GRBs, \citealt{Bykov1996, Xu2017, Summerlin2012}), the feedback heating in clusters of galaxies (\citealt{Guo2008, Brunetti2014, Zweibel2018}), driving Galactic winds (\citealt{Wiener2017, Krumholz2020}), and the confinement and re-acceleration of CRs in the Galaxy (\citealt{Chandran2000, Yan2002}).

In general, the most classical acceleration mechanisms are considered as the second- and first-order Fermi processes (\citealt{Fermi1949,Bell1978, Blandford1978}), and the (turbulent) magnetic reconnection (\citealt{Sweet1958, Parker1957, Petschek1964}; LV99). Note that the second-order Fermi is also called stochastic acceleration process. This model, originally proposed by \cite{Fermi1949}, suggests that particles can statistically gain energy through collisions with interstellar clouds, which is similar to the reflection of particles due to magnetic mirror effects. Since MHD waves in the turbulence provide the motion of scattering centers, particles can be continuously scattered to advance (\citealt{Melrose1980}).   

As is well known, scattering is considered an essential process for CR acceleration. For instance, the scattering of CRs back into the shock is a necessary component of the first-order Fermi acceleration (see \citealt{Longair2011}). At the same time, stochastic acceleration by turbulence is entirely based on the scattering process. Extensive numerical and analytical studies have been performed to understand the interactions of CRs with MHD turbulence, such as the scattering and diffusion processes (e.g., \citealt{Yan2002, Yan2004, Xu2013, Lazarian2021}). When the pitch-angle\footnote{Defining $\theta$ to be the pitch-angle between the particle velocity and the mean magnetic field, we have $\mu={\rm cos}\theta={\rm cos}90^\circ = 0$.} approaches to $90^\circ$, the scattering will vanish, and the mean free path becomes infinite (\citealt{Fisk1974}) since the particles are resonated by a very large wave-number $k$. This $90^\circ$ problem is one of the most famous concerns in quasi-linear theory's (QLT) predictions (\citealt{Jokipii1966}). The root reason causing the $90^\circ$ problem is the assumption of unperturbed trajectories in QLT. In order to avoid this problem, \cite{Yan2008} extends the QLT to nonlinear theory (termed NLT) by introducing finite resonance widths. 

At present, there are a lot of simulation works focused on exploring the diffusion and scattering of particles in MHD turbulence. One part is associated with acceleration (e.g., \citealt{Michalek1999, Beresnyak2011, Cohet2016}), while another with pure scattering and diffusion without acceleration (e.g., \citealt{Xu2013, Maiti2022, Hu2022}). The works involving acceleration, are not comprehensive and have their limitations in application to a complex turbulent environment. Specifically, \cite{Michalek1999} only focused on a single turbulence regime, i.e., the cold plasma limit, where fast and slow (magnetosonic) waves degenerate to fast mode waves. \cite{Beresnyak2011} was based on the case of incompressible turbulence using MHD plus test particle simulations. In addition, \cite{Cohet2016} focused on the sub-Alfv\'enic turbulence and explored the influence of driving ways on particle acceleration, where they found solenoidal forcing results do not match with QLT. Recently, \cite{Mertsch2020} provided a more comprehensive overview of test particle simulations of CRs, which is based on QLT, its extensions, and the state-of-the-art in the test particle simulations mainly from the perspective of synthetic data simulation.

In the earlier studies, it was claimed that the Alfv\'en mode is inefficient for scattering and acceleration of CRs (\citealt{Chandran2000, Yan2002}) due to its anisotropic properties (\citealt{Cho2002, Cho2003}, hereafter CL02 and CL03), while the fast mode is a major source of CRs scattering in the interstellar and intracluster media (\citealt{Yan2004, Brunetti2007}; see also \citealt{Michalek1999} for the case of fast mode domination) due to its isotropic properties (CL02 and CL03). However, \cite{Zhang2021} demonstrated from the perspective of particle spectral distribution that the contribution of Alfv\'en mode to particle accelerations cannot be ignored, which even plays a dominant role in particle acceleration at the late stage of acceleration. In the case of incompressible turbulence, it has been claimed that the role of pseudo-Alfv\'en mode, i.e., slow mode, is crucial for the scattering of particles (\citealt{Beresnyak2011, Xu2020}).

Furthermore, \cite{Demidem2020} claimed that the contributions of three modes to acceleration are comparable to within an order of magnitude in the relativistic MHD turbulence with Monte Carlo simulation of the test particle and synthetic data. It has been claimed that in the relativistic case, the properties of the three modes are similar to that of the non-relativistic one, except for the fast mode significantly coupling Alfv\'en mode (\citealt{Takamoto2016, Takamoto2017}). Recently, \cite{Yuen2023} found that the pure Alfv\'en mode can be decomposed into an anomalous compressible component, which implies that the Alfv\'en mode may affect the particle transport. From the perspective of analytical research, in the framework of the modern understanding of MHD turbulence theory, \cite{Cho2006} studied particle acceleration arising from fast and slow modes in both gaseous pressure-dominated (high $\beta$) and magnetic pressure-dominated (low $\beta$) cases. They predicted that the fast mode can accelerate particles more efficiently than the slow mode, and whether slow and fast modes dominate the acceleration of particles depends on the rate of spatial diffusion of particles. These interesting analytical predictions need to be tested and confirmed numerically. This motivates us to perform the current work by exploring various turbulence regimes.

One of the purposes of the current work is to study the diffusion and scattering behavior of particles being accelerated, and a second aim is to explore which plasma mode dominates the accelerated particle’s transport in various turbulence regimes. We investigate how energetic particles are accelerated in MHD turbulence, and test which of the analytical predictions in \cite{Cho2006} are consistent with our numerical results in a wide range of turbulence regimes that correspond to different astrophysical environments. Specifically, we want to know whether the fast mode dominates the acceleration, diffusion, and scattering process of particles exactly as the theoretical prediction, and whether the effect of the Alfv\'en mode is so weak as to be negligible.

This paper is organized as follows. In Section \ref{sec:theor}, we briefly introduce the theoretical description of MHD turbulence and diffusive properties of CRs. We perform test particle simulation and give our initial simulation set-up in Section \ref{sec:numersim}. We present the numerical results, which contain the Mach numbers (Alfv\'enic Mach number $M_{\rm A}$, and sonic Mach number ${M_{\rm s}}$) effect in four different turbulence regimes and the contributions of the three modes (wavelike isotropic fast mode, Alfv\'en, and slow Goldreich-Sridhar type modes) to particle acceleration and diffusion in Sections \ref{sec:numre.energ} and \ref{sec:numre.plas}, respectively. Discussion and summary are provided in Sections \ref{sec:discu} and \ref{sec:summa}, respectively.

\section{Theoretical descriptions}\label{sec:theor}
\subsection{MHD Turbulence Theory}\label{sec:theor.MHD}
The propagation of CRs is determined by their interactions with environmental turbulence. MHD turbulence theory has gone through a long period of development from pioneer works by Kolmogorov (\citeyear{Kolmogorov1941}; henceforth K41) and Iroshnikov \& Kraichnan (\citeyear{Iroshnikov1963, Kraichnan1965}; henceforth IK65) to the modern MHD turbulence theory by Goldreich \& Sridhar (\citeyear{Goldreich1995}; hereafter GS95) (see also \citealt{Schekochihin2020} for a recent review). By dimensional analysis, assuming that the energy is injected at a constant velocity ${\bm v}$ at large scales, K41 obtained the energy spectrum of $E(k) \propto k^{-5/3}$ for the pure fluid turbulence, called famous Kolmogorov spectrum. A few decades later, the IK65 spectrum, i.e., $E(k) \propto k^{-3/2}$, was proposed for magnetized fluid turbulence. Its shortcoming is that it incompletely predicted an isotropic cascade of MHD turbulence. Later, GS95 suggested a scale-dependent anisotropy, i.e., $k_{\parallel} \approx k_{\perp}^{2/3} L^{-1/3}$ for incompressible MHD turbulence, where $L$ is the outer scale of the turbulence, and $k_{\parallel}$ and $k_{\perp}$ is the parallel and perpendicular components with regard to the local magnetic field of the wave vector $k$, respectively. 

The modern theoretical understanding of MHD turbulence is based on the GS95 model. In the framework of eddy motions, LV99 and \cite{Lazarian2006} generalized incompressible MHD turbulence to compressible one in $M_{\rm A}> 1$ and $M_{\rm A}<1$, respectively. In the case of $M_{\rm A} < 1$, with a sub-Alfv\'enic velocity driving turbulence at injection scale $L_{\rm inj}$, the weak turbulence cascade spans the range from $L_{\rm inj}$ to $l_{\rm tr}$, where $l_{\rm tr} = L_{\rm inj}M_{\rm A}^{2}$ is the transition scale at which we have turbulence velocity $v_{l}$ equal to Alfv\'en speed $V_{\rm A}$, while the strong turbulence cascade is in the range of $[l_{\rm dis}, l_{\rm tr}]$, where $l_{\rm dis}$ is the dissipation scale. And in this range, we have the relationship of
\begin{equation}
l_{\parallel}\approx L_{\rm inj}^{1/3}l_{\perp}^{2/3}M_{\rm A}^{-4/3} \label{anis}
\end{equation}
associated with the parallel scale $l_{\parallel}$ and transversal one $l_{\perp}$, which suggests that eddies are stretched along the local magnetic field. Note that the original GS95 relation of $l_{\parallel} \propto l_{\perp}^{2/3}$ can be recovered by setting $M_{\rm A}=1$ in Equation (\ref{anis}). 

As for $M_{\rm A} > 1$, i.e., the super-Alfv\'enic turbulence, the cascade starting from the injection scale $L_{\rm inj}$ has a hydrodynamic Kolmogorov property, because of the marginal influence of a weak magnetic field on the cascade process. When the cascade is decreased to $l_{\rm tr} = L_{\rm inj}M_{\rm A}^{-3}$ (\citealt{Lazarian2006}), it enters a regime of strong turbulence, from which to the dissipation scale it again exhibits the GS95 anisotropy characteristics, with the scale relation of $l_{\parallel}\approx L_{\rm inj}^{1/3}l_{\perp}^{2/3}M_{\rm A}^{-1}$ and the velocity-scale relation of 
$v_{\rm l} = V_{\rm L}(l_\perp/L_{\rm inj})^{1/3}$, where $V_{\rm L}$ is the turbulence injection velocity at the scale $L_{\rm inj}$.

Furthermore, the analytical solution of the MHD dispersion equations can give three solutions (e.g., \citealt{Beresnyak2019}) that correspond to Alfv\'en, fast, and slow modes as numerically confirmed by CL02. Among the three modes, Alfv\'en mode is incompressible, while slow and fast modes are compressible and are called magnetoacoustic waves/modes. The latter two have scale-dependent anisotropic properties of $l_{\parallel} \propto l_{\perp}^{2/3}$ in the local frame of the magnetic field (LV99; \citealt{Cho2000, Maron2001}). Slow mode, passively mixed by Alfv\'en mode (\citealt{Lithwick2001}), has the same anisotropic scaling as Alfv\'en mode. Therefore, Alfv\'en and slow modes present the K41 spectrum of $E(k_{\perp}) \propto k_{\perp}^{-5/3}$. Differently, fast mode has isotropic properties of $l_{\parallel} \propto l_{\perp}$, reappearing the IK65 spectrum of $E(k) \propto k^{-3/2}$.

\subsection{Physical Quantities That Characterize CR Propagation}\label{sec:theor.physical quantities}
In general, the propagation and acceleration of CRs can be described by the Fokker-Planck equation coefficients (see the book from \citealt{Schlickeiser2002}). Here, we mainly focus on the coefficients relevant to our following simulations, such as the pitch-angle, spatial, and momentum diffusion coefficients. From a theoretical point of view, the diffusion of CRs results from the resonant (gyroresonance) and nonresonant (transit time damping, i.e., TTD) interaction of CRs with MHD turbulence. However, from the perspective of numerical simulation, it is currently difficult to distinguish the contribution of individual components.

As for the pitch-angle scattering, the pitch-angle diffusion coefficient is defined by
\begin{equation}
D_{\mu\mu} = \frac{\langle({\mu}-{\mu}_{\rm 0})^{2}\rangle}{2t}, \label{eq:dmumu}
\end{equation}
where the square brackets indicate an average over the ensemble of particles at the integration time $t$. The pitch angle $\theta$ is the angle between the particle velocity vector and the local magnetic field direction, whose cosine value is $\mu={\rm cos} \theta$ at the time $t$, while $\mu_0$ is at the initial moment $t_0$. Here, pitch angle $\theta$ range is from 0 to $90^{\circ}$, which corresponds to both $\mu$ and $\mu_0$ from 1 to 0. In the QLT, the mirror resonance has a sharp peak at $90^{\circ}$ due to discrete Landau resonant condition, resulting in the disappearance of the diffusion coefficient (\citealt{Goldstein1976,Felice2001}). This is the infamous 90-degree problem. To avoid the $90^{\circ}$ problem, QLT was extended to NLT by taking the magnetic mirroring effect into account on large scales (\citealt{Yan2008}). In addition, when introducing the bouncing of CRs, this problem of quasilinear gyroresonant can be alleviated (see \citealt{Lazarian2021} for the details).

Using the pitch-angle diffusion coefficient, we can determine the parallel mean free path of the particles by substituting $D_{\mu\mu}$ into (\citealt{Earl1974})
\begin{equation}
\frac{\lambda_{\parallel}}{L_{\rm inj}} = {\frac{3}{4}}\int_{0}^{1}d{\mu}{\frac{{u}(1-{\mu}^2)^2}{D_{\mu\mu}L_{\rm inj}}}, \label{eq:lambda1}
\end{equation}
where $u$ is the velocity of particles. Alternatively, the mean free path $\lambda_{\parallel}$ of particles can be calculated by 
\begin{equation}
\lambda_{\parallel} = \frac{3D_{\parallel}}{u}.
\label{eq:lambda2}
\end{equation}
Here, $D_{\parallel}$ is a parallel diffusion coefficient (\citealt{Giacalone1999})
\begin{equation}
D_{\parallel} = \frac{\langle(\tilde{x}-\tilde{x_0})^2\rangle}{2t}, \label{eq:dpara}
\end{equation}
where $(\tilde{x}-\tilde{x_0})$ is the spatial separation in the local magnetic field directions. Numerically, it is not difficult to test the fact that Equations (\ref{eq:lambda2}) and (\ref{eq:lambda1}) give similar results (see also \citealt{Maiti2022} for their testing). Similarly, we can also obtain the perpendicular diffusion coefficient
\begin{equation}
D_{\perp} = \frac{\langle(\tilde{y}-\tilde{y_0})^2\rangle}{2t}, \label{eq:dper}
\end{equation}
where $(\tilde{y}-\tilde{y_0})$ represents the spatial separation perpendicular to the local magnetic fields.

Assuming that the particles move in terms of a random walk in momentum space, we can determine the momentum diffusion coefficient
\begin{equation}
D_{pp} = \frac{\langle(\Delta p)^2\rangle}{2\Delta t} \label{eq:dpp}
\end{equation}
averaged over the ensemble of particles, where $\Delta p$ is the amount of change in particle momentum in the time interval $\Delta t$. In this work, we use Equation (\ref{eq:dpp}) to characterize the particle diffusion behavior in the acceleration processes. 

\section{Numerical simulation}\label{sec:numersim}
\subsection{Simulation of MHD Turbulence}\label{sec:numersim.MHD}
The third-order-accurate hybrid, essentially non-oscillatory code is adopted to solve the ideal MHD equations describing MHD turbulence as follows: 
\begin{equation}
{\partial \rho }/{\partial t} + \nabla \cdot (\rho {\bm v})=0, \label{eq:8}
\end{equation}
\begin{equation}
\rho[\partial {\bm v} /{\partial t} + ({\bm v}\cdot \nabla) {\bm v}] +  \nabla p_{\rm g}
        - {\bm J} \times {\bm B}/4\pi ={\bm f}, \label{eq:9}
\end{equation}
\begin{equation}
{\partial {\bm B}}/{\partial t} -\nabla \times ({\bm v} \times{\bm B})=0,\label{eq:10}
\end{equation}
\begin{equation}
\nabla \cdot {\bm B}=0,\label{eq:11}
\end{equation}
where $p_{\rm g} = c_{\rm s}^2\rho$ is the gas pressure, $c_{\rm s}$ and $\rho$ represent the sonic speed and density, respectively, $t$ is the time of fluid evolution, ${\bm J} = {\nabla} \times {\bm B}$ is the current density, and ${\bm f}$ is a random solenoidal driving force on large scale (small wavenumber $k \simeq 2.5$ in wavenumber space) in our simulation. The other parameters have their usual meaning. Our simulation is performed in a periodic box at the length of $L=2\pi$, setting a non-zero mean magnetic field strength along the $x$-axis direction. When statistical steady-state is reached, we output primitive physical quantities such as the magnetic field, velocity, and density. To characterize each simulation, we calculate Alfv{\'e}nic number by $M_{\rm A}=\langle \frac{\vert \bm v \vert}{\bm v_{\rm A}} \rangle$, and sonic Mach number by $M_{\rm s}=\langle \frac{\vert \bm v \vert}{c_{\rm s}} \rangle$, where $v_{\rm A}= \frac{B_0}{4\pi\sqrt{\rho}}$ is the Alfv{\'e}n speed. The resulting values are listed in Table \ref{table_1}, where each model corresponds to different turbulence regimes.

\begin{deluxetable*}{cccccccccc}
\tabletypesize{\scriptsize}
%\rotate
\tablewidth{170mm}
\tablehead{
\colhead{Models} & \colhead{$B_0$} & \colhead{$M_{\rm A}$}
   & \colhead{$M_{\rm s}$} & \colhead{$\beta$}
   & \colhead{$\langle {B_{\rm A}}^2\rangle:\langle{B_{\rm F}}^2\rangle:\langle{B_{\rm S}}^2\rangle$}
   & \colhead{$\langle{V_{\rm A}}^2\rangle:\langle{V_{\rm F}}^2\rangle:\langle{V_{\rm S}}^2\rangle$}
   & \colhead{$\langle{B_{\rm A}}^2\rangle:\langle{V_{\rm A}}^2\rangle$}
   & \colhead{$\langle{B_{\rm F}}^2\rangle:\langle{V_{\rm F}}^2\rangle$}
   & \colhead{$\langle{B_{\rm S}}^2\rangle:\langle{V_{\rm S}}^2\rangle$}
}
\startdata
A1   & 1.0    & 0.65   & 0.48   & 3.67    &  0.95 : 1.11 : 7.94    &  4.15 : 2.28 : 3.57     &  1.37 : 8.63     &  2.53 : 7.47      &  6.08 : 3.92 \\
A2   & 1.0    & 0.55   & 4.46   & 0.03    &  1.73 : 7.89 : 0.38    &  5.42 : 2.75 : 1.83     &  1.31 : 8.69     &  5.75 : 4.25     &  0.89 : 9.11  \\
A3   & 0.1    & 1.72   & 0.45   & 29.21   &  2.30 : 0.01 : 7.69    &  4.09 : 1.87 : 4.04     &  9.59 : 0.41     &  1.91 : 8.09     &  9.88 : 0.12  \\
A4   & 0.1    & 1.69   & 3.11   & 0.59    &  2.16 : 0.52 : 7.32    &  3.86 : 1.82 : 4.32     &  9.72 : 0.28     &  9.47 : 0.53     &  9.91 : 0.09  \\
A5   & 0.1    & 0.50   & 9.92   & 0.01   &  1.71 : 8.21 : 0.08    &  5.42 : 2.65 : 1.93     &  1.27 : 8.73     &  5.88 : 4.12     &  0.19 : 9.81  \\
\enddata
\caption{Data cubes with numerical resolution of $512^3$ for different turbulence regimes. $M_{\rm A}$ and $M_{\rm s}$ are the Alfv\'enic and sonic Mach numbers, respectively, and $\beta = p_{\rm gas}/p_{\rm mag} = 2M_{\rm A}^2/M_{\rm s}^2$ is the plasma parameter. $\langle {B_{i}}^2\rangle$ and $\langle {V_{i}}^2\rangle$ denote
the magnetic and kinetic energies corresponding to Alfv\'en ($i=\rm A$), fast ($i=\rm F$), and slow ($i=\rm S$) modes, respectively.
}
\label{table_1}
\end{deluxetable*}

Numerically, the compressible MHD turbulence was first decomposed into Alfv\'en, slow and fast modes by a Fourier transformation (CL02; CL03). The limitation of this method is that it only applies to the global reference frame and can only deal with the problems of $M_{\rm A} < 1$. Later,  \cite{Kowal2010} improved this separation technique by introducing a discrete wavelet transformation before the Fourier separation, extending it to the $M_{\rm A} > 1$ case. The displacement vectors of the slow, fast, and Alfv\'en modes are defined by their unit vectors
\begin{equation}
\hat{\zeta}_{\rm s} \varpropto (-1 + \alpha - \sqrt{D})(k_{\parallel} \hat {\bm k}_{\parallel})
+(1 + \alpha- \sqrt{D})(k_{\perp} \hat {\bm k}_{\perp}), \label{eq:12}
\end{equation}

\begin{equation}
\hat{\zeta}_{\rm f} \varpropto (-1 + \alpha + \sqrt{D})(k_{\parallel} \hat {\bm k}_{\parallel})
+(1 + \alpha + \sqrt{D})(k_{\perp} \hat {\bm k}_{\perp}), \label{eq:13}
\end{equation}

\begin{equation}
\hat{\zeta}_{\rm A} \varpropto \hat{\bm k}_{\perp} \times \hat{\bm k}_{\parallel},\label{eq:14}
\end{equation}
with $D = (1+\alpha)^2-4\alpha{\cos^{2}\phi}$ and $\alpha = c_{\rm s}^2/V_{\rm A}^2$, where $\phi$ is an angle between ${\bm k}$ and ${\bm B}_{0}$. After numerically determining the $\hat{\zeta}_{\rm s}$, $\hat{\zeta}_{\rm f}$ and $\hat{\zeta}_{\rm A}$, we can project the total magnetic field and velocity into these three unit vector directions and obtain information about the magnetic and velocity fields of each mode. To maintain consistency in the operation of all data, we use the wavelet decomposition in this work.

\subsection{Method of Test Particle}\label{sec:numersim.testp}
In the electromagnetic field, the kinetic equation that the motion of a charged particle satisfies is
\begin{equation}
\frac{d}{dt}({\gamma}m{\bm u}) = q({\bm E}+{\bm u} \times {\bm B}), \label{eq:kin1}
\end{equation}
where ${\gamma} \equiv 1/\sqrt{1-u^2/c^2}$ is the Lorentz factor of the particle, $m$ and $q$ are the mass and charge of the particle. In this work, we take into consideration the acceleration process resulting from magnetic field $\bm B$ and its inductive field ${\bm E} = -{\bm v} \times {\bm B}$, without considering the resistivity of the electric field. Therefore, Equation (\ref{eq:kin1}) can be rewritten as
\begin{equation}
\frac{d}{dt}({\gamma}m{\bm u}) = q[({\bm u}-{\bm v}) \times {\bm B}]. \label{eq:kin2}
\end{equation}
By integrating this equation, we can trace particle's trajectories using the classic 4th order explicit Runge-Kutta (RK4) method with adaptive time step.\footnote{Compared the results from different integrators, we obtained consistent results with that of the RK4 method. See footnote 4 of \cite{Zhang2023} for more details.} We can recover the local values of the plasma velocity $\bm v$ and magnetic field $\bm B$ at each step of the integration by using linear interpolation with the discontinuity detector. 

When analyzing the numerical output, we calculate the direction of the local magnetic fields by \citep{Cho2000}
\begin{equation}
{{\bm B}_l} = \frac{{\bm B}({{\bm r}_1})+{\bm B}({{\bm r}_2})}{2}. \label{eq:bl}
\end{equation}
And then we can define $D_{\parallel}$ as the parallel direction of $\hat{\bm x} ={{\bm B}_l}/|{{\bm B}_l}|$ and $D_{\perp}$ as its perpendicular direction.

\begin{figure*}
\centering
\includegraphics[width=1.8\columnwidth,height=0.6\textheight]{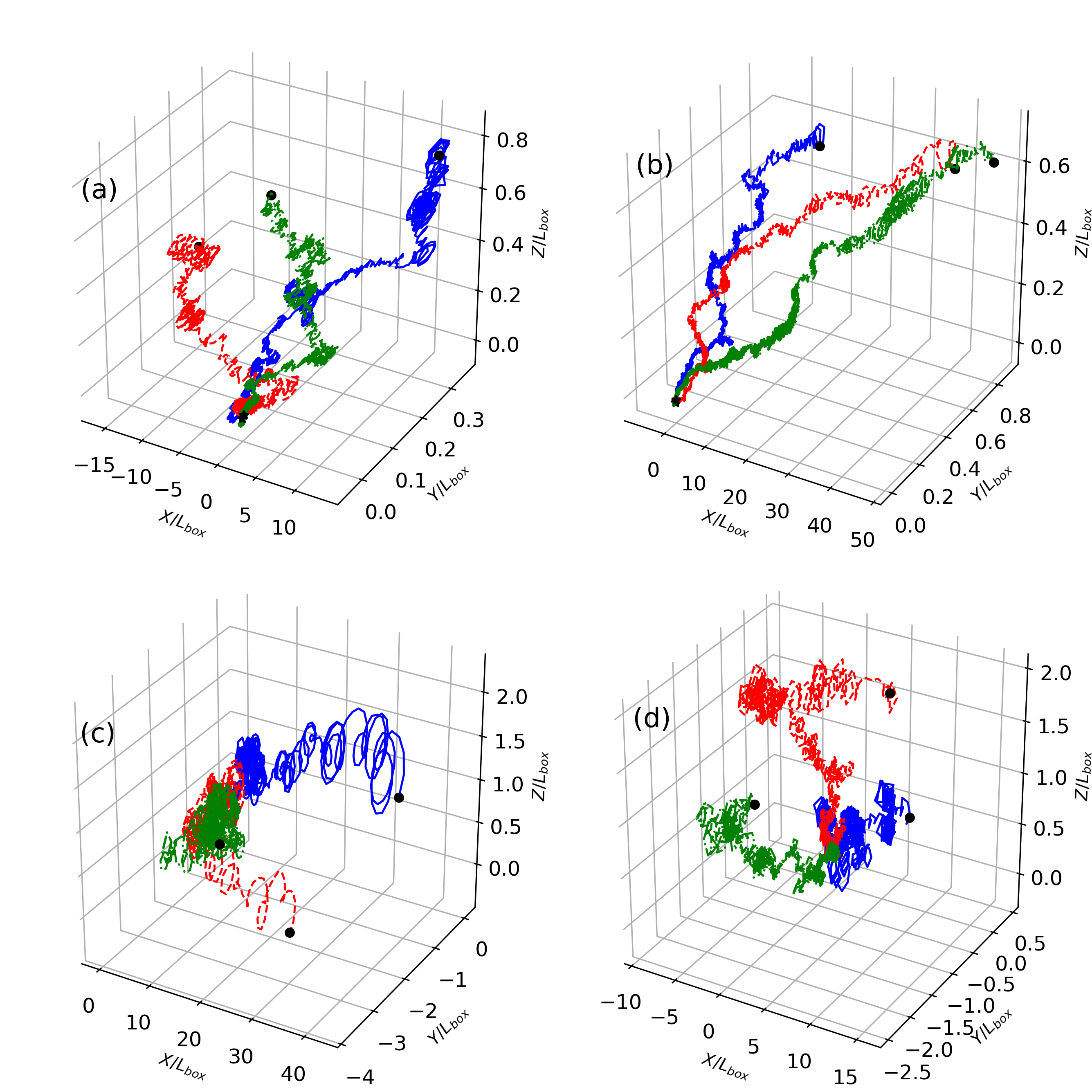}
\caption{The 3D trajectories of three test particles selected, arising from sub-Alfv\'enic \& subsonic (panel a), sub-Alfv\'enic \& supersonic (panel b), super-Alfv\'enic \& subsonic (panel c), and super-Alfv\'enic \& supersonic (panel d) turbulence regimes. The star and circle markers refer to their initial and final locations in the 3D space, respectively. $L_{\rm box} = {\rm 512}$ is the box scale.
}
\label{fig:01trajectory}
\end{figure*}

\section{Numerical Results: the influence of turbulence properties on particle energization}\label{sec:numre.energ}
In this section, we provide numerical results of particles' energization processes related to different turbulence regimes, based on data cubes from simulations of MHD turbulence listed in Table \ref{table_1}. Specifically, our results include the trajectory and gyroradius of particles, their momentum, spatial and pitch-angle diffusion in four turbulence regimes such as sub-Alfv\'enic \& subsonic, sub-Alfv\'enic \& supersonic, super-Alfv\'enic \& subsonic, and super-Alfv\'enic \& supersonic.

\subsection{Trajectory of Particles}\label{sec:numre.energ.traje}
At the beginning of the simulation, we instantly inject $10^4$ test particles with a Maxwell-type spectrum randomly distributed into the whole box space of the simulation. Throughout this paper, we set the temperature of test particles to be $10^6$ K, the plasma velocity (i.e., the fluid velocity) to be $5\times 10^7\ {\rm m/s}=0.1667\ c$, and magnetic field strength to be $5\ \mu{\rm Gs}$. 

\begin{figure*}[t]
\centering
\includegraphics[width=0.95\textwidth,height=0.45\textheight]{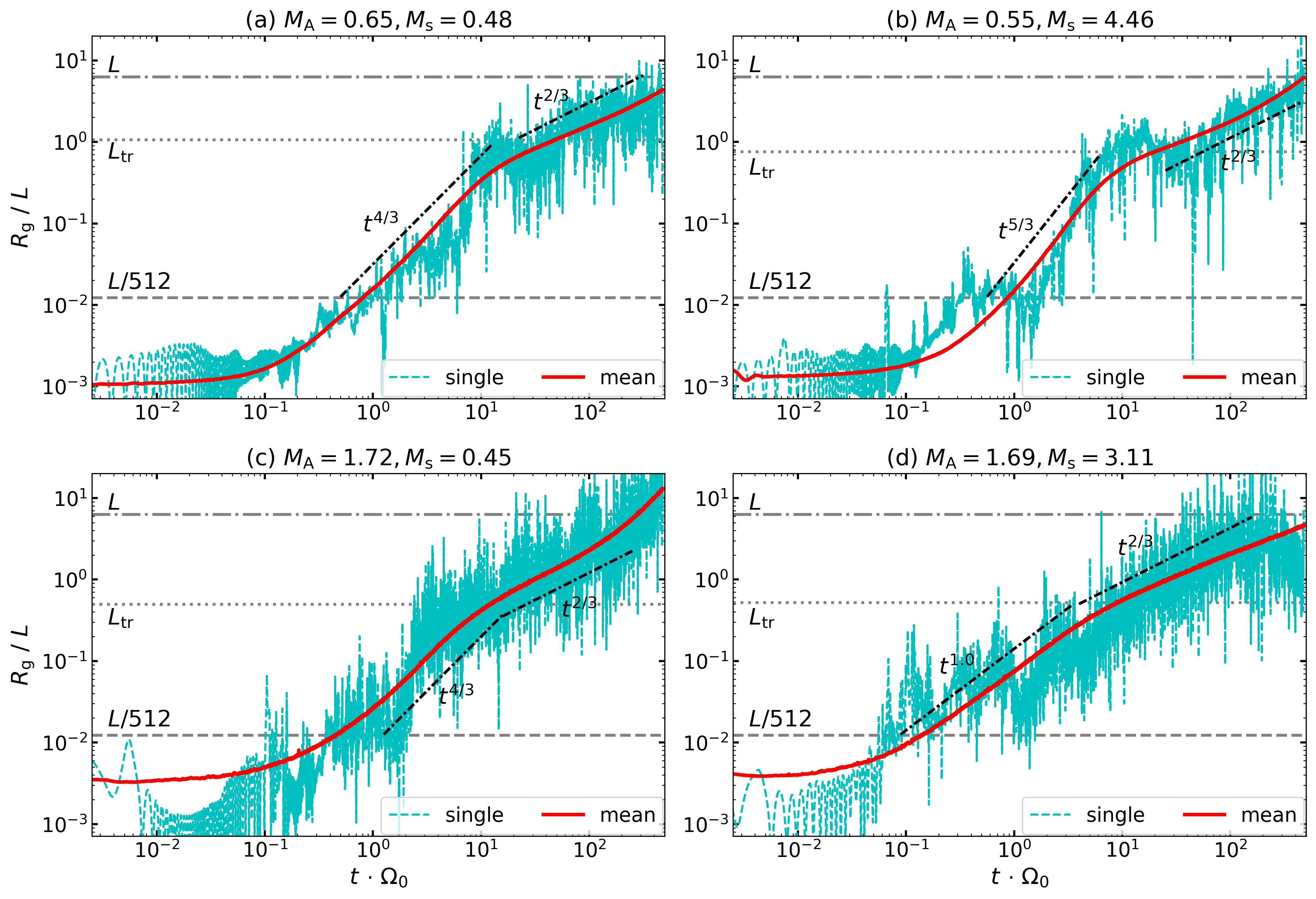}
\caption{The time evolution of gyroradius of a single particle is compared with the behavior of an ensemble of $10^{4}$ test particles interacting with four different turbulence scenarios. The horizontal dashed, dotted and dash-dotted lines show the grid, transition and box scales, respectively. $L$ and $\Omega_0$ are the box length and initial gyrofrequency, respectively.
    }
    \label{fig:02basicpara1}
\end{figure*}

Figure \ref{fig:01trajectory} plots the trajectories of the 1000th, 5000th, and 7000th test particles. As shown, the particles experience diffusive motion in the simulation space. We see that the particles cross a large spatial scale in the $x$-axis direction, which is because, in our simulation of MHD turbulence, the mean magnetic field is set along this direction. Compared with panels (a) and (b), we see also that the particles in panels (c) and (d) span a larger space in three coordinate axis directions. Given that the setting of the magnetic field strength of $5.0\ \rm \mu Gs$, we have the dimensionalized mean magnetic field strength of $B_0=5.0\ \rm \mu Gs$ for sub-Alfv\'enic turbulence (panels a and b) and of $B_0=0.5\ \rm \mu Gs$ for super-Alfv\'enic turbulence (panels c and d). According to the formula of gyro motion of a charged particle,
\begin{equation}
{R_{\rm g}} = \frac{p_{\perp}}{\lvert q \rvert {B}} = \frac{{\gamma}m{v}_{\perp}}{\lvert q \rvert {B}}, \label{eq:rg}
\end{equation}
where $p_\perp = {\gamma}mv_\perp$ is the perpendicular momentum of a particle at the perpendicular velocity $v_\perp$ with respect to the local magnetic field, $B$, it is not difficult to understand particles' motion behavior. For a fixed $p_\perp$, the larger the $B$ value is, the smaller the $R_{\rm g}$. As a result, particles have a smaller gyroradius, resulting in a smaller spatial extension (upper panels). Another interesting point is that in the case of super-Alfv\'enic \& subsonic turbulence (lower-left panel), particles have the largest gyroradii at the final acceleration time (see also the lower-left panel of Figure \ref{fig:02basicpara1}), gaining the maximum kinetic energy of $E_{\rm max} = 715.73 \ m_{\rm p}{\rm c}^2$. This implies that a strong mean magnetic field does not necessarily result in the maximum possible energy.

Except for tracking the trajectory of a single particle, we also check particle's drift motion by observing the evolution of particle velocities and displacement over time (not shown here). We find that the perpendicular velocity with respect to the local mean magnetic field is always greater than the parallel one. As for the time evolution of the displacement, we find that the perpendicular displacement is much larger than the parallel one in the early stage of the evolution, which indicates that magnetic field gradient drift may dominate the acceleration. However, in the late stage, the larger parallel displacement indicates the domination of turbulence cascade interaction, which is consistent with the results from diffusion analysis (see Section \ref{sec:numre.energ.spati}).

\subsection{Momentum Diffusion}\label{sec:numre.energ.momen}
Figure \ref{fig:02basicpara1} plots the evolution of the gyroradius over time for four turbulence regimes, where the red solid lines represent the averaged gyroradius of all particles and the cyan dashed lines represent the gyroradius distribution of a single particle. As seen, the evolution of the gyroradius over time presents an overall consistency in the different turbulence regimes. Based on the length scales, we divide each evolution of $R_{\rm g}$ into three main stages. In the first stage ($R_{\rm g}<L/512$), the gyroradius $R_{\rm g}$ of particles slowly increases with the time after a plateau. This indicates that at the start of the simulation, the acceleration of the particles is inefficient, which implies that the particles cannot interact with turbulence waves (scattering centers) efficiently. The acceleration phenomenon of $R_{\rm g}$ less than one grid size ($L/512$) may arise from the non-resonant interaction between the particles and fluid, the gradient and curvature drifts of magnetic fields.

In the second stage ($L/512< R_{\rm g}<L_{\rm tr}$), we see that $R_{\rm g}$ shows a power-law relationship of $t^{4/3}$ for the subsonic turbulence, while for the supersonic turbulence, $R_{\rm g}$ shows the power law of $t^{5/3}$ and $t^{1.0}$ in the sub-Alfv\'enic and super-Alfv\'enic regimes, respectively. In this strong turbulence regime we are mostly interested in, the acceleration efficiency of particle acceleration has been significantly improved, with a distribution ranging from $R_{\rm g}\propto t^{1.0}$ to $R_{\rm g}\propto t^{5/3}$. It arises from quasi-resonant interactions of particles with different scale eddies, that is, the interaction between fluid and magnetic fields causes the induced electric field to accelerate particles. In the third stage ($R_{\rm g}> L_{\rm tr}$), the time evolution of $R_{\rm g}$ for all four regimes shows a power law of $R_{\rm g}\propto t^{2/3}$ in this weak turbulence regime. This demonstrates that the efficiency of particle interaction with turbulence is decreased.

\begin{figure*}[t]
\centering
\includegraphics[width=0.48\textwidth,height=0.7\textheight]{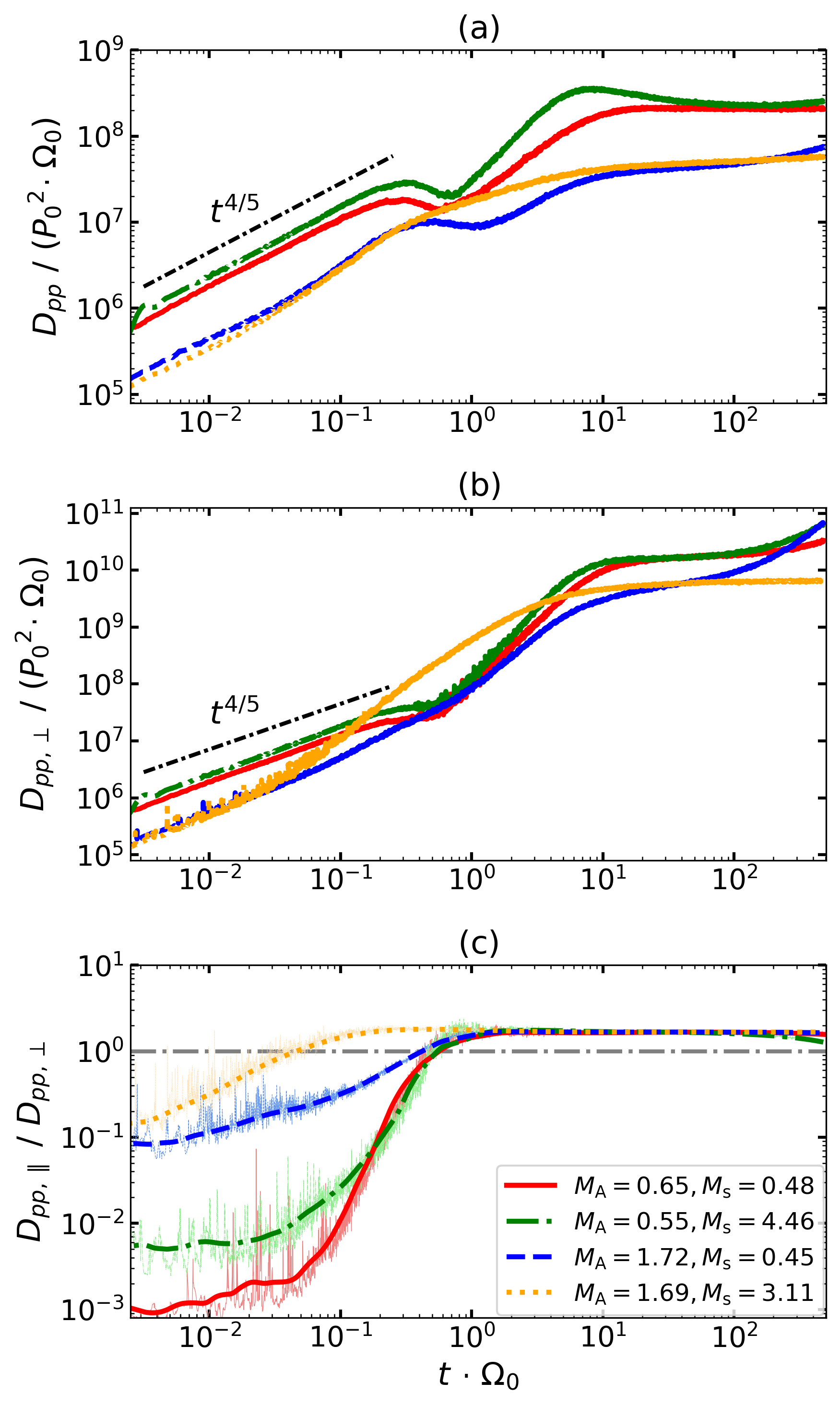} 
\includegraphics[width=0.48\textwidth,height=0.7\textheight]{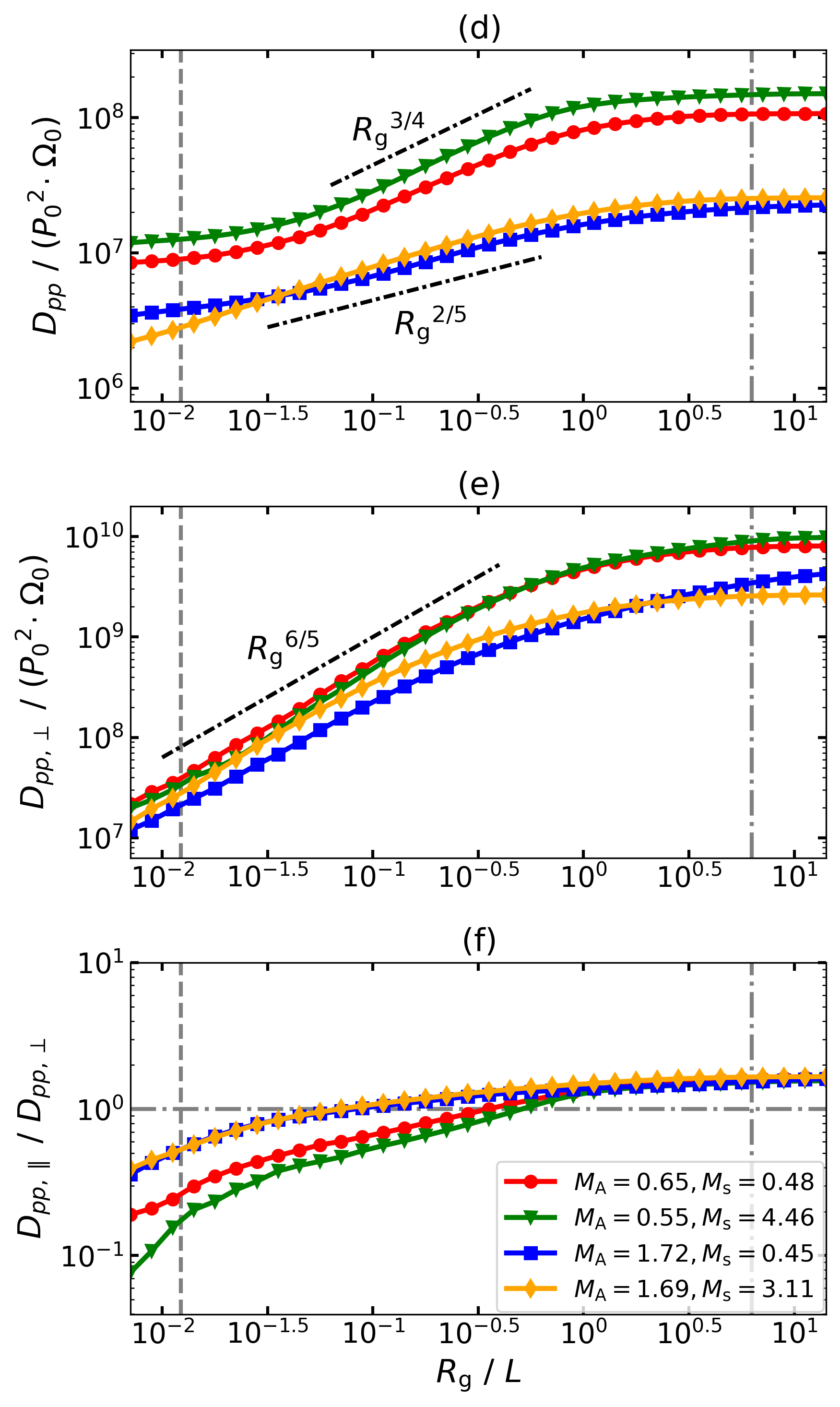}
\caption{The momentum diffusion coefficient as a function of the evolution time (left column) and the gyroradius of particles accelerated (right column). The total momentum diffusion coefficient, its perpendicular component, and the ratio of parallel to perpendicular component are plotted in the upper, middle, and lower rows, respectively. The vertical dashed and dash-dotted lines plotted in the right column present the grid and box sizes, respectively, while the horizontal dash-dotted line shows $D_{pp, \parallel}/D_{pp, \perp}=1$. $L$, $\Omega_0$, and $P_0$ are the box length, initial gyrofrequency, and momentum, respectively.
    }
    \label{fig:03dpp1}
\end{figure*}

Figure \ref{fig:03dpp1} shows the evolution of the momentum diffusion coefficient over time (left column) and gyroradius (right column), including the total momentum diffusion $D_{pp}$ (upper row), its perpendicular component $D_{pp, \perp}$ with respect to the local magnetic field (middle row), and the ratio of parallel to perpendicular components $D_{pp, \parallel}/D_{pp, \perp}$ (lower row).\footnote{Note that $D_{pp}$ is considered as a function of time $t$ instead of $\Delta t$. In fact, we find that the slope relationship from $\langle(\Delta p)^2\rangle$ vs. $\Delta t$ is consistent with that from $D_{pp}$ vs. $t$.} As shown in panel (a), before $1/{\Omega_0}$, corresponding to the first stage of $R_{\rm g}< L/512$ mentioned above, the total $D_{pp}$ over time presents a power-law relationship of $D_{pp}\propto t^{4/5}$. At $t \simeq 1/{\Omega_0}$, the evolution of $D_{pp}$ over time presents a trough (seen also in \cite{Demidem2020}) except for the super-Alfv\'enic and supersonic one (orange-dotted line), which indicates that the turbulence interaction tends to suppress the particle diffusion in the dissipation region. When entering the strong turbulence regime, the increasing rate of $D_{pp}$ for sub-Alfv\'enic turbulence gets higher with a steeper distribution. In these two stages, the particle undergoes the superdiffusion in the momentum space\footnote{$D_{pp}\propto t^{a_{p}-1}$, when $a_{p} < 1$, it is called subdiffusion; when $a_{p} = 1$, it is called normal diffusion; when $a_{p} > 1$, it is called superdiffusion (\citealt{Ostrowski1997, Sioulas2020}).}. When the gyroradius reaches the transition scale, i.e., enters the weak turbulence regime, the evolution of $D_{pp}$ follows a plateau phase, which is a significant feature of the second-order Fermi process (\citealt{Pezzi2022, Liang2023}) and means that the particle experiences the normal diffusion.

To explore the relationship between $D_{pp}$ and the gyroradius $R_{\rm g}$, we divide $R_{\rm g}$ from $10^{-3}/L$ to $10^{2}/L$ into 50 bins in the logarithmic bin, then pick out the particles in each bin to calculate their diffusion coefficients.\footnote{After dividing $R_{\rm g}$ into 30, 50, 100, and 500 bins, we find that the results are mostly stable at 30 bins, so we show the results for 50 bins in this paper.} In the box length range from $R_{\rm g}=L/512$ to $L$ that we are interested in, we plot in panel (d) the momentum diffusion coefficient as a function of $R_{\rm g}$. As shown, it approximates to $D_{pp}\propto {R_{\rm g}}^{3/4}$ in the strong turbulence regime for sub-Alfv\'enic turbulence, while $D_{pp}\propto {R_{\rm g}}^{2/5}$ for super-Alfv\'enic turbulence. This demonstrates that momentum diffusion in sub-Alfv\'enic turbulence is faster than that in super-Alfv\'enic turbulence. This should be the fact that in the momentum space, the strong magnetic fields improve particle diffusion. For both cases of sub- and super-Alfv\'enic turbulence, $D_{pp}$ shows a plateau after $R_{\rm g}> L_{\rm tr}$, i.e., after $10/ \Omega_0$; see also panel (a). Furthermore, we also explore the evolution of $D_{pp}$ with momentum $p$, which shows a similar power-law relationship with that of gyroradius, i.e. $D_{pp}\propto p^{3/4}$ for sub-Alfv\'enic turbulence, and $D_{pp}\propto p^{2/5}$ for super-Alfv\'enic one. Note that \cite{Cho2006} predicted $D_{pp}\sim (\Delta p)^2/\Delta t \sim p^2(\nabla \cdot \bm{v_l})^2 \Delta t$, from which we can reasonably speculate $\nabla \cdot \bm{v_l} \propto p^{5/8}/\Delta t$ and $\propto p^{4/5}/\Delta t$.

In the middle panels of Figure \ref{fig:03dpp1}, we plot the perpendicular component $D_{pp,\perp}$ over the integrated time and gyroradius (by binning the particles into different values of gyroradius). Its overall behavior is similar to $D_{pp}$, except for some small differences. In particular, we see a good power-law relation of $D_{pp, \perp}\propto t^{4/5}$ before $1 /{\Omega_0}$ and $D_{pp, \perp}\propto {R_{\rm g}}^{6/5}$ in the strong turbulence regimes for four turbulence scenarios. Moreover, the ratio of $D_{pp, \parallel}$ and $D_{pp, \perp}$ is shown in the lower panels, from which we can see that at the early stage of evolution $D_{pp, \perp}$ is even one order of magnitude larger than $D_{pp, \parallel}$ for the sub-Alfv\'enic \& super-Alfv\'enic turbulence regime. This implies that at small $R_{\rm g}$ the perpendicular gradient drift of the magnetic field dominates particle diffusion processes by $\bm{v}_{\rm grad}=\gamma m v_{\rm \perp}^2 (\bm{B}\times \nabla \bm{B})/2q^2B^2$. For the time evolution of $D_{pp}$ (panel c), after $t\sim 1 /{\Omega_0}$, the diffusion is dominated by the parallel momentum. For its gyroradius evolution (panel f), in the case of super-Alfv\'enic turbulence, when $R_{\rm g} > L/512$, $D_{pp, \parallel}/D_{pp, \perp}$ grows slowly and enters the stage dominated by parallel momentum, and in the sub-Alfv\'enic case, when $R_{\rm g} > L_{\rm tr}$ the parallel momentum dominates the momentum diffusion. This means that the diffusion is dominated by the parallel momentum at large $R_{\rm g}$.

\subsection{Spatial Diffusion and Pitch-Angle Scattering}\label{sec:numre.energ.spati}
\begin{figure}
\centering
\includegraphics[width=1.0\columnwidth,height=0.45\textheight]{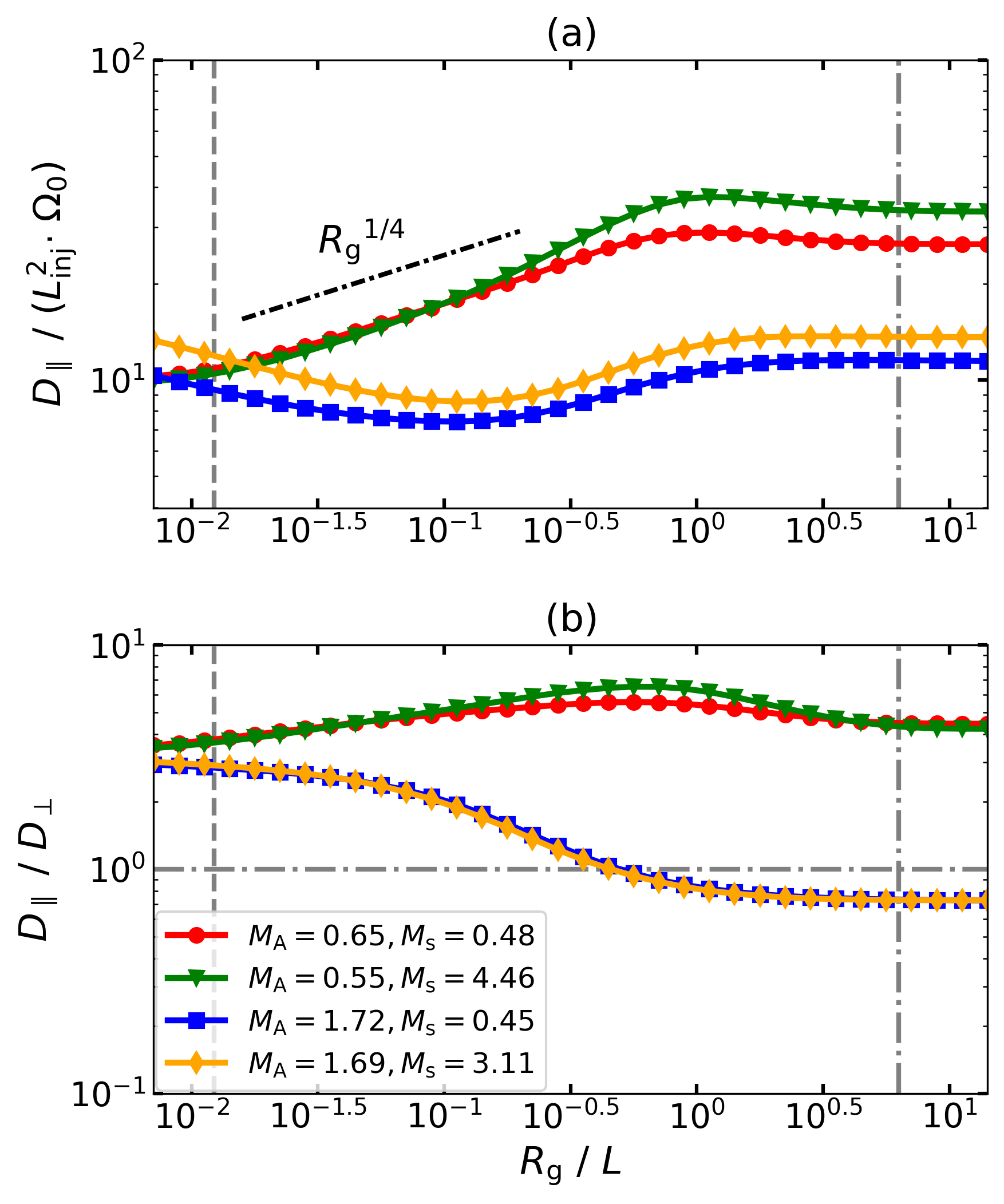}
\caption{The evolution of parallel diffusion coefficient (panel a), $D_{\parallel}$, and the ratio of $D_{\parallel}$ to perpendicular diffusion coefficient $D_{\perp}$ (panel b) with the gyroradius. The other descriptions are the same as those of Figure \ref{fig:03dpp1}.
}
\label{fig:04drr1}
\end{figure}

We bin particles into different values of gyroradius, then present the spatial diffusion coefficient as a function of the gyroradius in Figure \ref{fig:04drr1}, including the parallel diffusion $D_{\parallel}$ (panel a) and the ratio of parallel and perpendicular components (panel b) for four turbulence regimes. In panel (a), $D_{\parallel}$ for sub-Alfv\'enic turbulence shows a power-law relationship of $\propto R_{\rm g}^{1/4}$ and a plateau before and after $R_{\rm g} \sim 1 / L$, respectively. This indicates that the spatial diffusion of particles first increases in the strong turbulence regime, i.e., scales smaller than $L_{\rm tr}$, and then enters a plateau stage in the weak turbulence regime, i.e., scales greater than $L_{\rm tr}$. In the case of super-Alfv\'enic turbulence, $D_{\parallel}$ shows a slower slope at scales smaller than $L_{\rm tr}$, which implies that strong magnetic fields enhance the parallel spatial diffusion.

To gain insight into the detailed information of the spatial diffusion of the accelerated particles, we further show the ratio of parallel and perpendicular diffusion coefficients $D_{\parallel}/D_{\perp}$ in panel (b). In the case of $M_{\rm A} < 1$, the ratio is larger than 1 all the time, which means that $D_{\parallel}$ dominates the spatial diffusion in this case. As for the case of $M_{\rm A} > 1$, the distribution of the ratio can be divided into two stages: first, it is larger than 1 and slowly decreases, which means that the spatial diffusion of CRs is dominated by $D_{\parallel}$; second, after $R_{\rm g}\sim 0.3/ L$, the ratio is less than 1 and enters into the stage dominated by $D_{\perp}$. This indicates that in the strongly turbulent cascade process, super-Alfv\'enic turbulence tends to inhibit the parallel diffusion of the accelerated particles. Similarly, by analyzing the mean free path of the particles (see Equation \ref{eq:lambda2}), we also come to similar results.

\begin{figure}[t]
\centering
\includegraphics[width=0.48\textwidth,height=0.28\textheight]{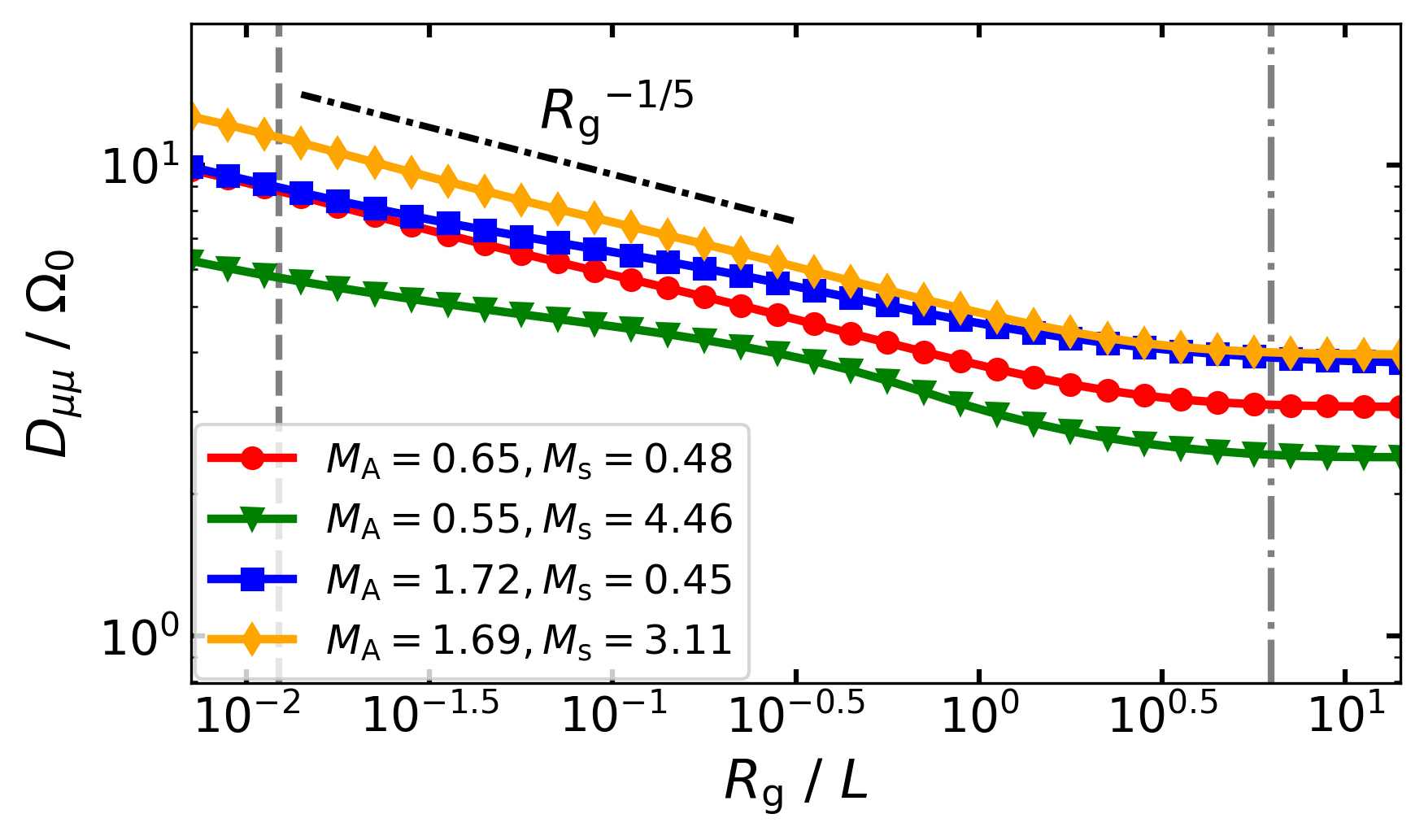}
\caption{The pitch-angle diffusion coefficient, $D_{\mu\mu}$, as a function of gyroradius at four turbulence regimes explored. The other descriptions are the same as those of Figure \ref{fig:03dpp1}.
}
\label{fig:05dmumu1}
\end{figure}

Next, with the aim of studying the pitch-angle scattering behavior of the accelerated particles, we bin particles into different values of $R_{\rm g}$ and show the evolution of $D_{\mu\mu}$ over $R_{\rm g}$ in Figure \ref{fig:05dmumu1}. With increasing $R_{\rm g}$, $D_{\mu\mu}$ can be roughly divided into two stages. Firstly, differences between these four turbulence regimes begin to emerge during the plateau stage of $R_{\rm g}$ vs. $t$ (see Figure \ref{fig:02basicpara1}). $D_{\mu\mu}$ of these four turbulence regimes presents a similar evolution with $R_{\rm g}$, which decays with a power-law relationship of $D_{\mu\mu}\propto {R_{\rm g}}^{-1/5}$, except for a slightly slower decay of the sub-Alfv\'enic and supersonic case (green triangle-down line). Secondly, when $R_{\rm g}$ of particles reaches the transition scale $L_{\rm tr}$, where the momentum and spatial diffusion reach a plateau, the decay rate of $D_{\mu\mu}$ slows down and gradually enters a plateau for these four regimes. Note that there are opposite evolutionary trends between $D_{\mu\mu}$ and $D_{pp}$ over $R_{\rm g}$. This suggests that when a particle is being accelerated, its pitch-angle scattering is suppressed. 

\section{Numerical Results: interaction of cosmic ray particles with plasma modes}\label{sec:numre.plas}
\subsection{Trajectory of Particles}\label{sec:numre.plas.traje}
\begin{figure*}
\centering
\includegraphics[width=1.8\columnwidth,height=0.6\textheight]{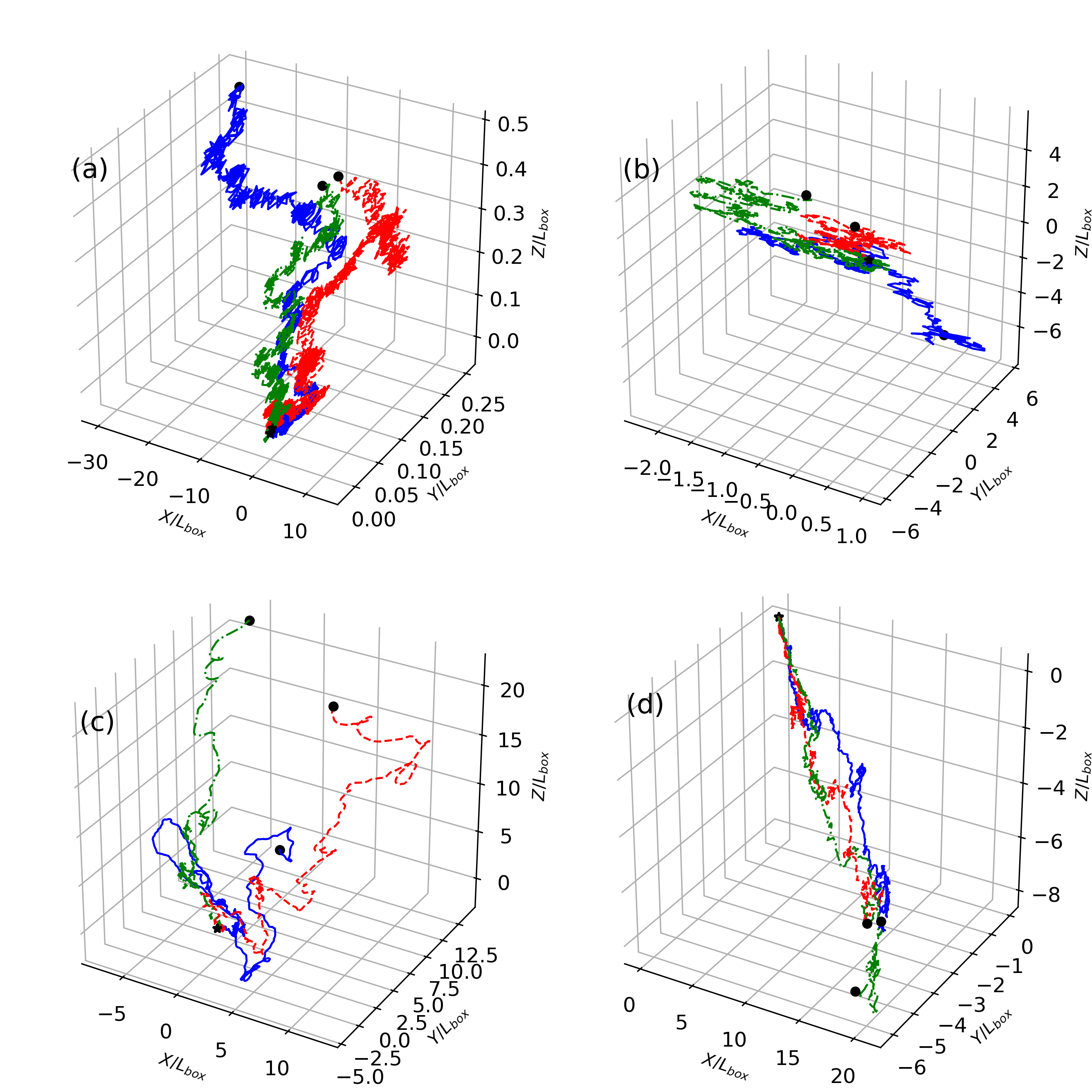}
\caption{The 3D trajectories of three test particles selected, arising from data A5 (panel a) listed in Table \ref{table_1} and its decomposed modes including the Alfv\'en (panel b), fast (panel c), and slow (panel d). The star and circle markers refer to their initial and final locations in the 3D space, respectively. $L_{\rm box} = 512$ is the box scale.
}
\label{fig:06trajectoryA5}
\end{figure*}

Based on data cubes A1 and A5 listed in Table \ref{table_1} from simulations of MHD turbulence, we provide numerical results of CRs interacting with decomposed plasma modes. Figure \ref{fig:06trajectoryA5} plots the trajectories of selected three particles from the pre-decomposed MHD turbulence (panel a) and its post-decomposed three modes: the Alfv\'en (panel b), fast (panel c), and slow (panel d) modes. The pre-decomposed MHD turbulence shows a similar diffusive motion as Figure \ref{fig:01trajectory}. As for the post-decomposed plasma modes, the trajectories for the fast mode are similar to that of pre-decomposed turbulence, and trajectories for the Alfv\'en mode are extended along the $x$-axis direction, which is consistent with our expectation due to its intrinsically polarized feature. Although the slow mode has similar trajectories with the Alfv\'en one, which should be from their same scale-dependent anisotropies (\citealt{Cho2002}), it experiences more extended spatial lengths. This phenomenon implies that the particle interaction with turbulence for fast mode is more effective, and the particle acceleration rate should be higher in the supersonic regime. In addition, we see that the spatial scales in the $x$-axis direction for slow and fast modes are slightly larger than that for Alfv\'en mode, which may mean that magnetosonic modes have a slightly higher diffusion efficiency than Alfv\'en mode. These inferences will be further confirmed below.

\subsection{Momentum Diffusion}\label{sec:numre.plas.momen}
\begin{figure*}[t]
\centering
\includegraphics[width=0.48\textwidth,height=0.5\textheight]{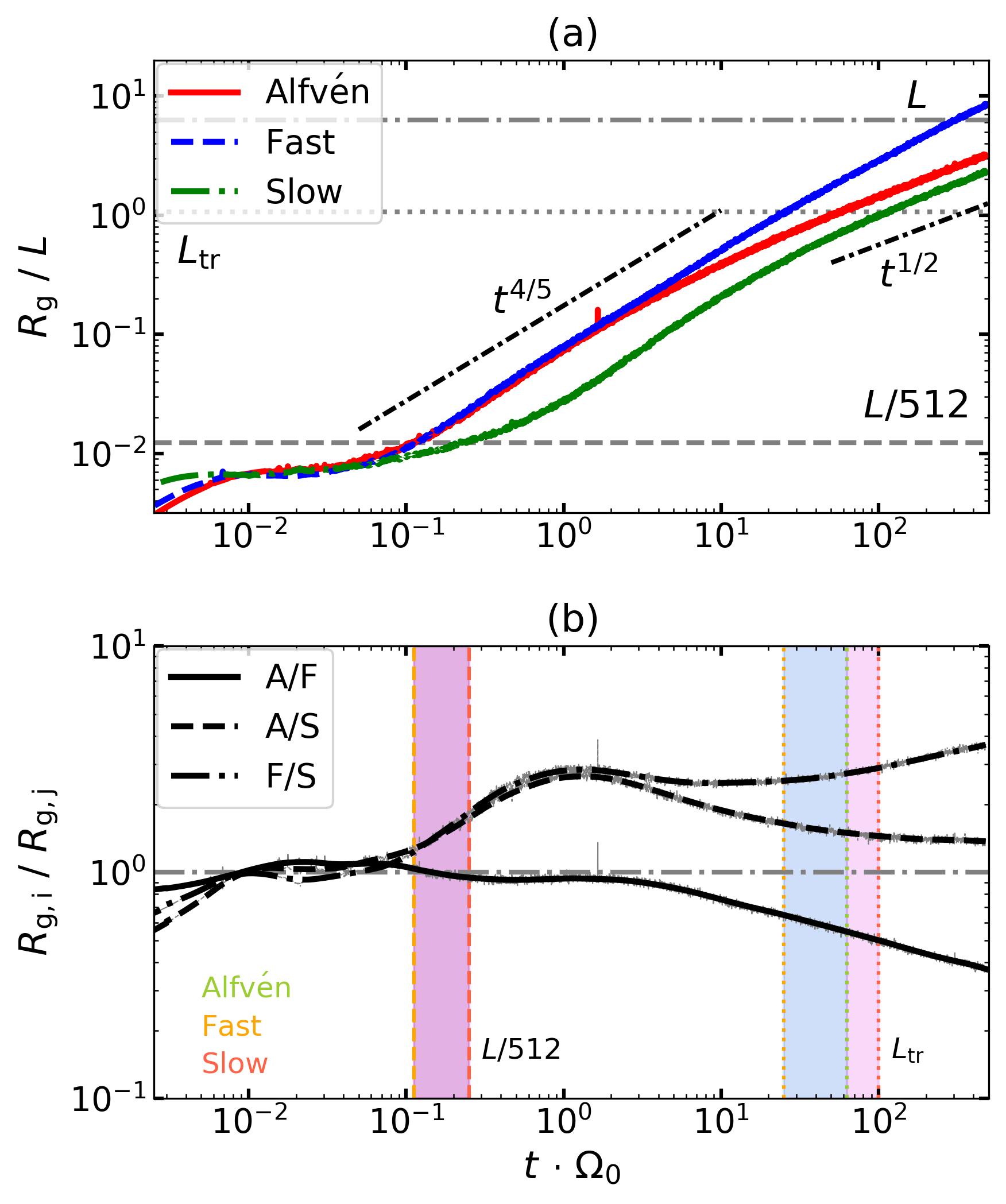} 
\includegraphics[width=0.48\textwidth,height=0.5\textheight]{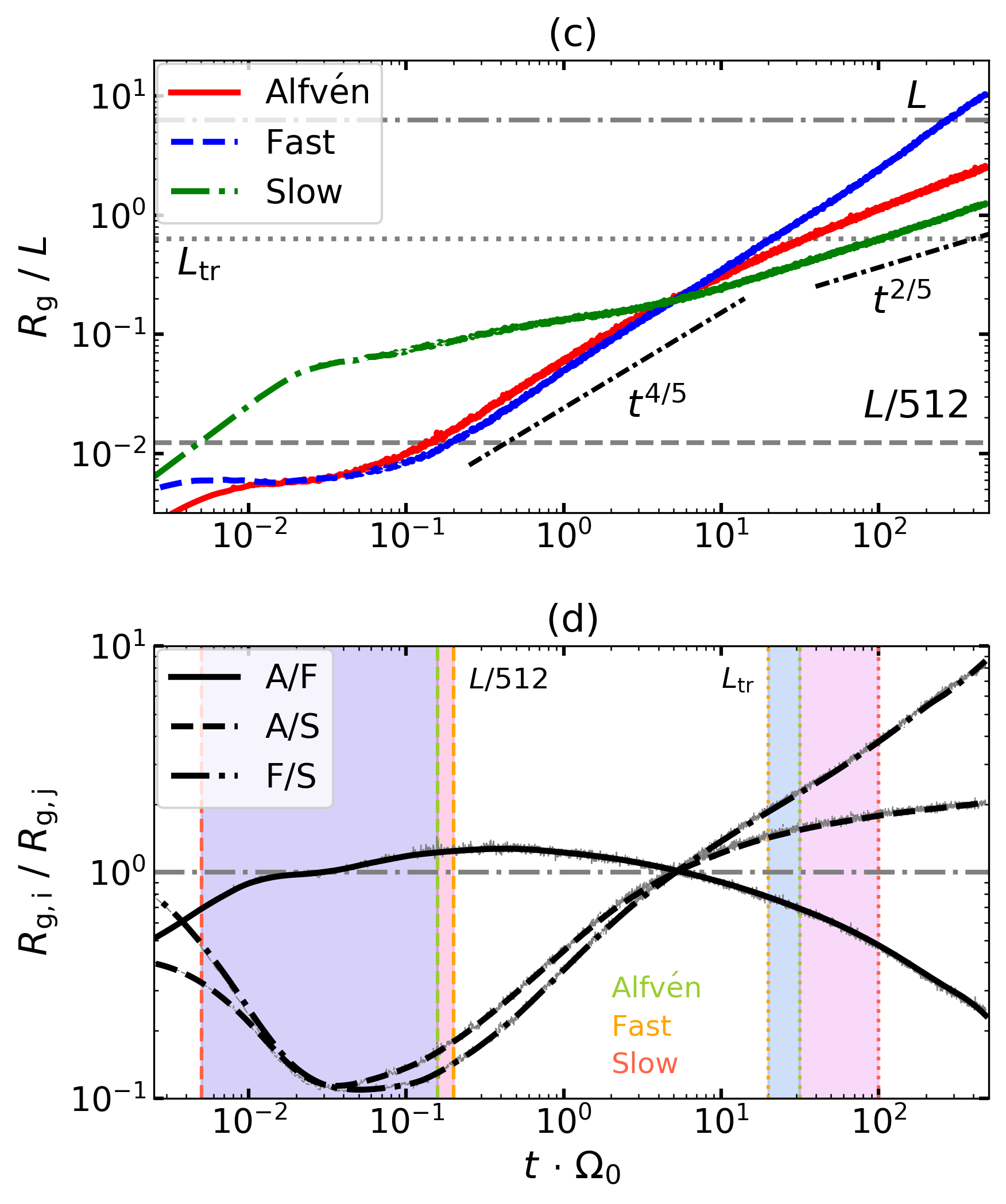}
\caption{The gyroradius of the accelerated particles as a function of the evolution time for decomposed three plasma modes (upper row) and the ratio of three modes (lower row). The left and right columns are based on A1 and A5 listed in Table \ref{table_1}, respectively. The subscript $i$ represents Alfv\'en and fast modes, while the subscript $j$ slow and fast modes. The vertical dashed and dotted lines represent the time that the gyroradius reaches the grid and transition scales, respectively, where the colors yellow-green, orange, and tomato present the Alfv\'en, fast, and slow modes, respectively. $L$ and $\Omega_0$ are the box length and initial gyrofrequency, respectively.
    }
    \label{fig:07basicpara2}
\end{figure*}

In Figure \ref{fig:07basicpara2}, we show the gyroradius of particles from the three modes and their ratios in the high and low $\beta$ regimes.\footnote{It should be noted that, when plotting the figures, we used the same transition scale related to the same $M_{\rm A}$ calculated from the full MHD data. This is only to approximate the corresponding scale relationship to explain the simulation results.} As shown in the left column for the high $\beta$ regime, we can see that the acceleration processes can be divided into three different stages by the length scales. In the first stage ($R_{\rm g}<L/512$), the gyroradii of these three modes present a plateau, and then slowly increase over time to exceed $L/512$. In the second stage ($L/512 <R_{\rm g}<L_{\rm tr}$), i.e., the strong turbulence regime, $R_{\rm g}$ of fast and Alfv\'en modes is very similar to each other presenting a power-law relationship of $t^{4/5}$, and larger than $R_{\rm g, S}$ at the same simulation time. This may be related to the fraction of magnetic energies listed in Table \ref{table_1}, that is, the larger the $B$ value, the $R_{\rm g}$ is smaller (see Equation (\ref{eq:rg})). In the third stage ($R_{\rm g}>L_{\rm tr}$), i.e., the weak turbulence regime, the fast mode remains the same power law as that in the second stage, while Alfv\'en and slow modes show a shallower power law of $R_{\rm g} \propto t^{1/2}$. Interestingly, the acceleration behavior of particles caused by three plasma modes is different from  that caused by overall MHD turbulence before being decomposed (see Figure \ref{fig:02basicpara1}). 

To explore their ratios, we highlight the timescale zone in panel (b) that the gyroradii of three modes reach the $L/512$ and $L_{\rm tr}$ scales, by filling in the area between two adjacent vertical lines at each scale for convenience. We can more clearly see their differences: during the plateau stage (i.e., the left region of the pink vertical bandwidth), $R_{\rm g, A} \simeq R_{\rm g, F} \simeq R_{\rm g, S}$; after that (i.e., the region between the pink and cornflower-blue vertical bandwidth), $R_{\rm g, S}$ is almost always smaller than $R_{\rm g, A}$ and $R_{\rm g, F}$; as for the weak turbulence regime (i.e., the right region of the violet vertical bandwidth), it is opposite that $R_{\rm g, F}$ is the largest one then followed by $R_{\rm g, A}$.

In the case of the low $\beta$ regime (right column), there is a distinctly different phenomenon for the slow mode before about $10^{0.5} /{\Omega_0}$ (during the strong turbulence regime). Slow wave starts to accelerate before the plateau stage of fast and Alfv\'en modes. After $10^{0.5} /{\Omega_0}$, $R_{\rm g}$ of the slow mode grows slowly and presents a power-law relationship of $R_{\rm g} \propto t^{2/5}$, compared with the high $\beta$ scenario. As seen, in the first two stages (the left region of the corn-flower-blue vertical bandwidth), the interaction caused by the slow mode dominates the acceleration process, while in the third stage (the right region of the corn-flower-blue vertical bandwidth), the acceleration efficiency of the fast mode is higher than that of the other two. In this comparison study, since Models A1 and A5 listed in Table \ref{table_1} have very close values of $M_{\rm A}$ = 0.65 and 0.5, respectively, the difference in $\beta$ is mainly caused by the difference in $M_{\rm s}$ (0.48 for A1 and 9.92 for A5) by the relationship of $\beta=2M_{\rm A}^2/M_{\rm s}^2$. Therefore, we find that the shock wave occurring in the supersonic case influences the particle's acceleration.

\begin{figure*}[t]
\centering
\includegraphics[width=0.48\textwidth,height=0.6\textheight]{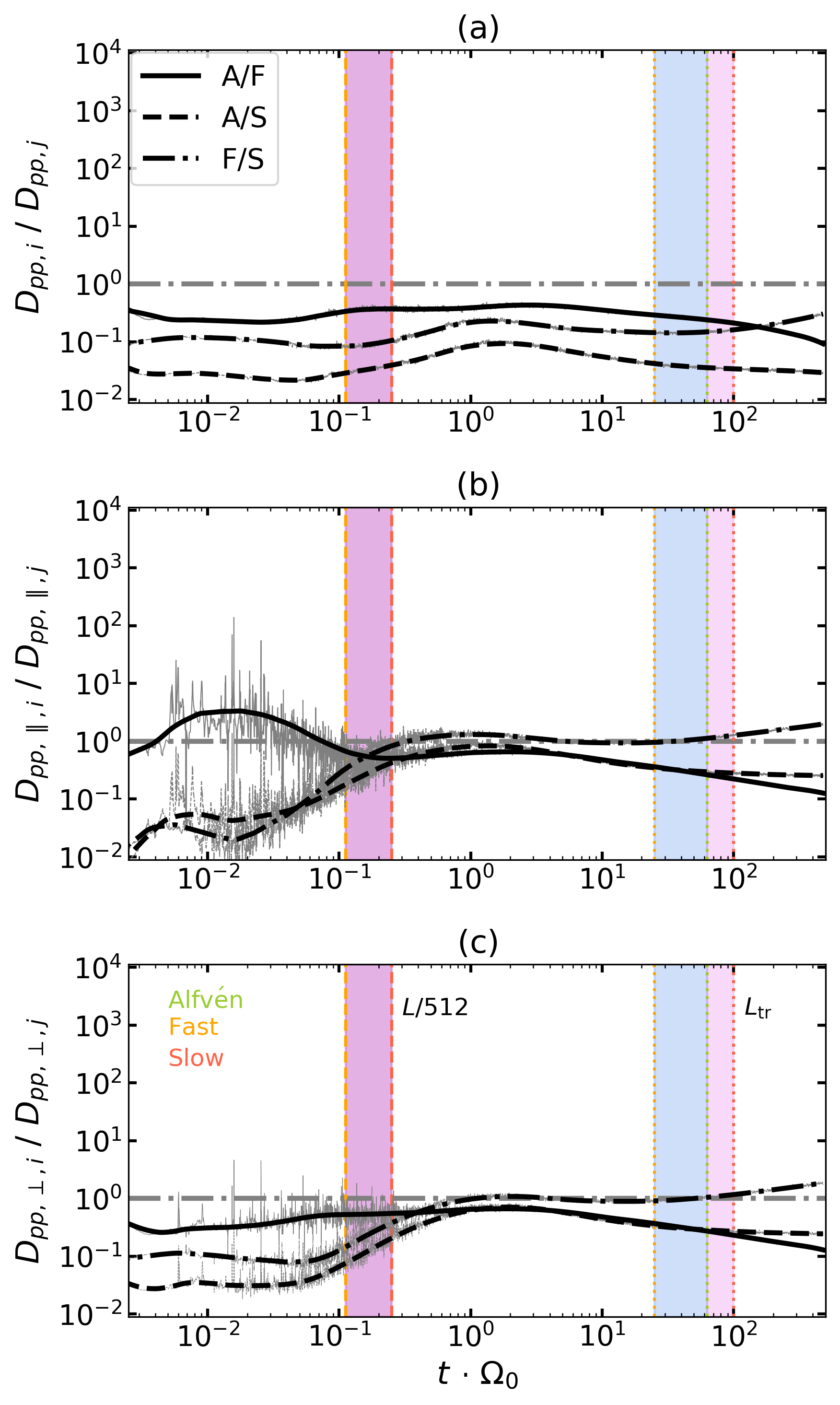}
\includegraphics[width=0.48\textwidth,height=0.6\textheight]{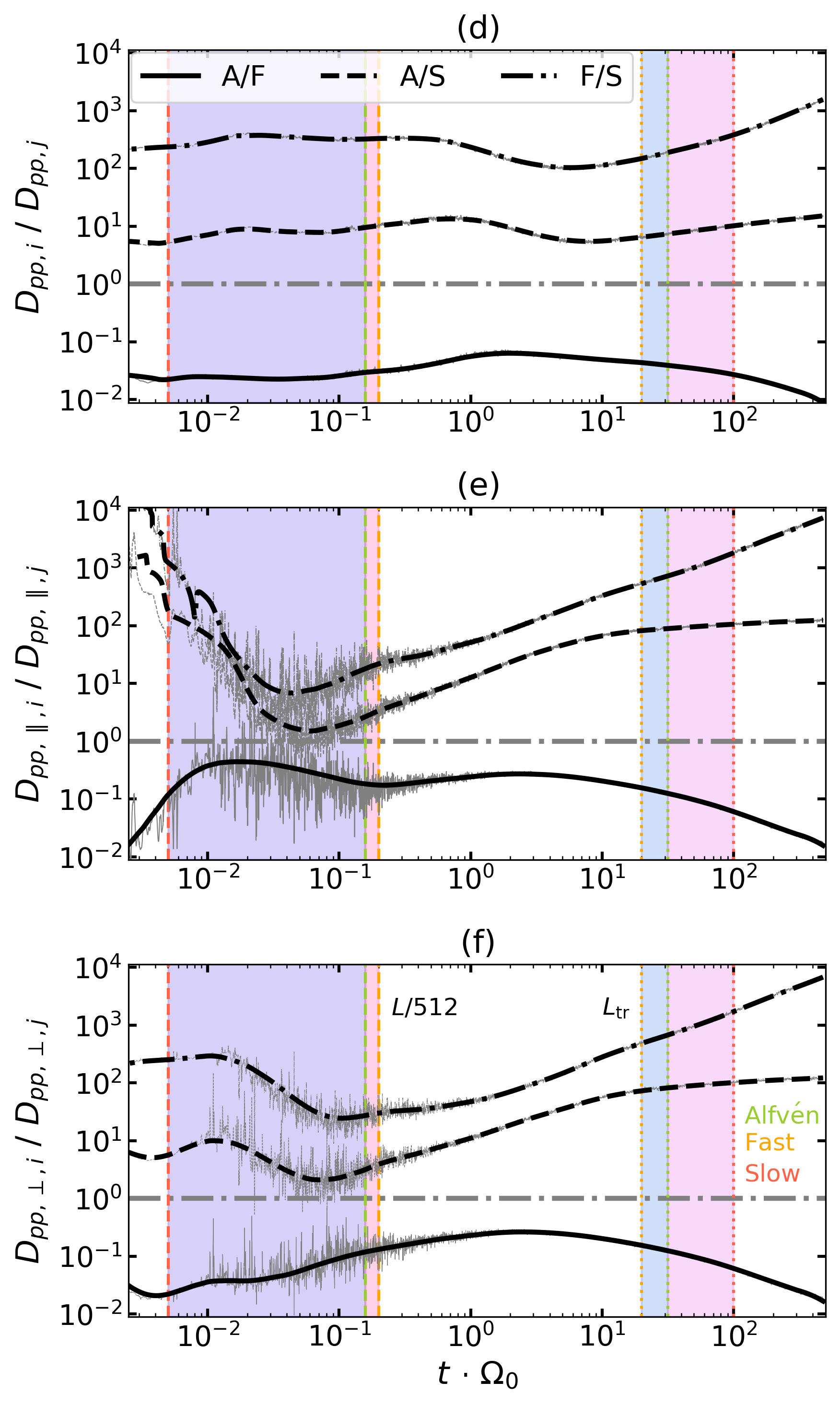}
\caption{The ratio of momentum diffusion coefficients between two modes as a function of the evolution time arising from total momentum (top row), and its parallel (middle row) and perpendicular (bottom row) components, based on A1 (left column) and A5 (right column) as listed in Table \ref{table_1}. The light curves with many fluctuations are original distributions, while the dark curves are a robust local weighted regression fitting. The other descriptions are the same as those of Figure \ref{fig:07basicpara2}.
}
\label{fig:08dpp2-afs}
\end{figure*}

To explore the effect of the plasma modes on the diffusion in momentum space, we show in Figure \ref{fig:08dpp2-afs} the ratio of momentum diffusion coefficients between two modes as a function of the time, for the total momentum (top row), its parallel $D_{pp, \parallel}$ (middle row), and perpendicular $D_{pp, \perp}$ (bottom row) components. In the high $\beta$ regime (left column), for the total momentum diffusion coefficient (panel a), the contribution of magnetosonic modes ($D_{pp, S}$ and $D_{pp, F}$) is greater than Alfv\'en mode ($D_{pp, A}$) during the whole evolution. As for $D_{pp, \parallel}$ (panel b) and $D_{pp, \perp}$ (panel c), it is basically the same as the total momentum diffusion coefficient, though $D_{pp, F} / D_{pp, S}$ is close to one after about $1 /{\Omega_0}$, corresponding to the last two stages of the time evolution of gyroradius. This means that in the high $\beta$ case, the fast mode is as crucial as the slow mode for the parallel and perpendicular components of momentum diffusion in the later stages of the particle acceleration.

In the low $\beta$ regime (right column), the contribution of the fast mode is almost always the largest one, followed by Alfv\'en mode, and slow mode is the smallest, including the total momentum diffusion coefficient $D_{pp}$, as well as its parallel and perpendicular components ($D_{pp, \parallel}$ and $D_{pp, \perp}$). In detail, the order of magnitude of the ratio of any two modes in low $\beta$ is slightly higher than that in high $\beta$. Especially for $D_{pp, \parallel}$, the fast mode is almost several orders of magnitude higher than the slow mode in the first stage.

In short, our one finding is that the fast mode in the high $\beta$ case dominates the particle acceleration, while in the low $\beta$ case, the fast and slow modes dominate the acceleration. Here, the dominance of the fast mode is in agreement with the earlier theoretical predictions (\citealt{Yan2002,Chandran2003,Cho2006}). Another finding is that our numerical results confirm the dominance of magnetosonic modes in the acceleration and momentum diffusion of CRs, which is consistent with the previous studies (\citealt{Yan2004, Lynn2014,Zhang2020}).

\subsection{Spatial Diffusion and Pitch-Angle Scattering}\label{sec:numre.plas.spati}
\begin{figure*}[t]
\centering
\includegraphics[width=0.48\textwidth,height=0.5\textheight]{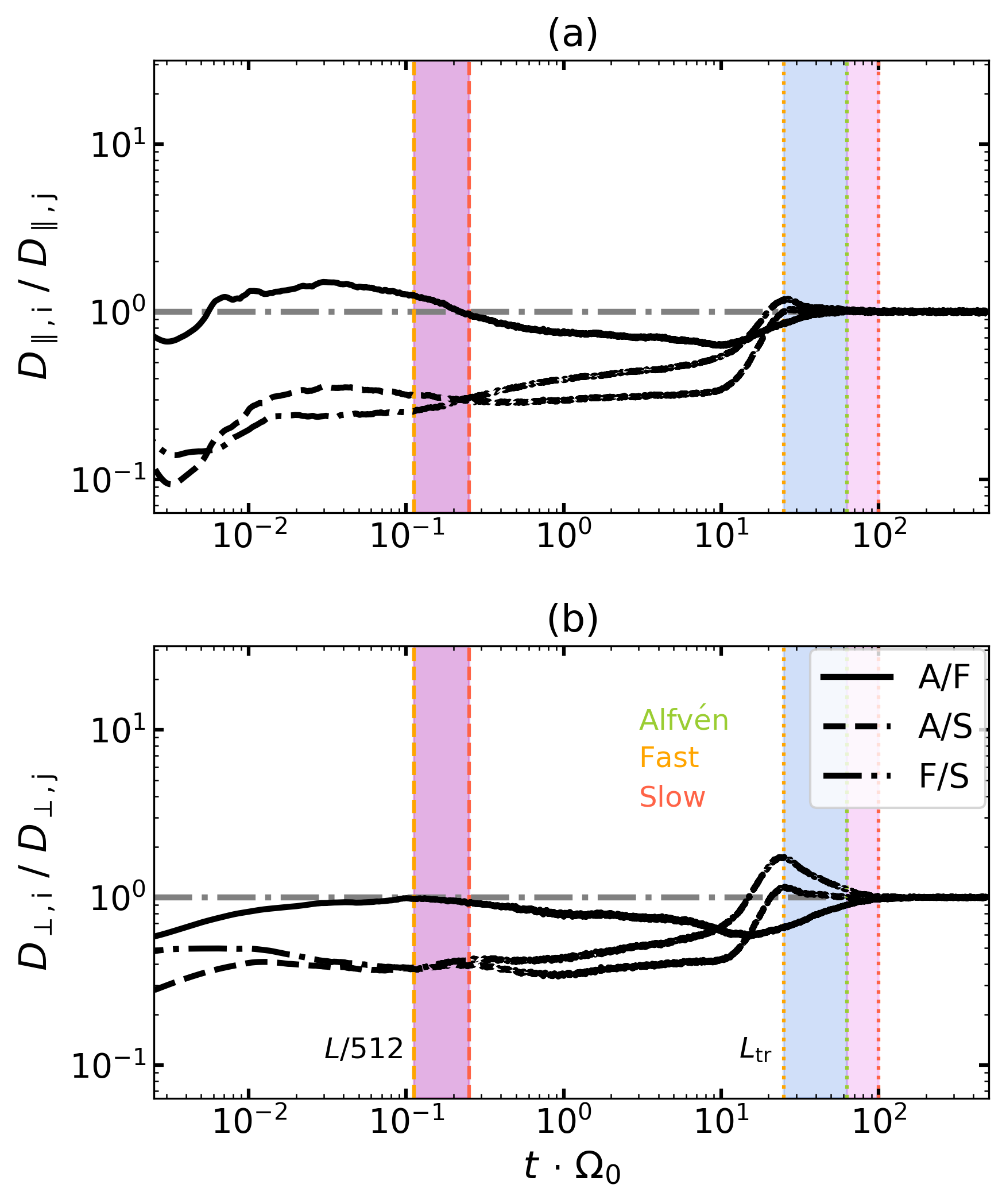}
\includegraphics[width=0.48\textwidth,height=0.5\textheight]{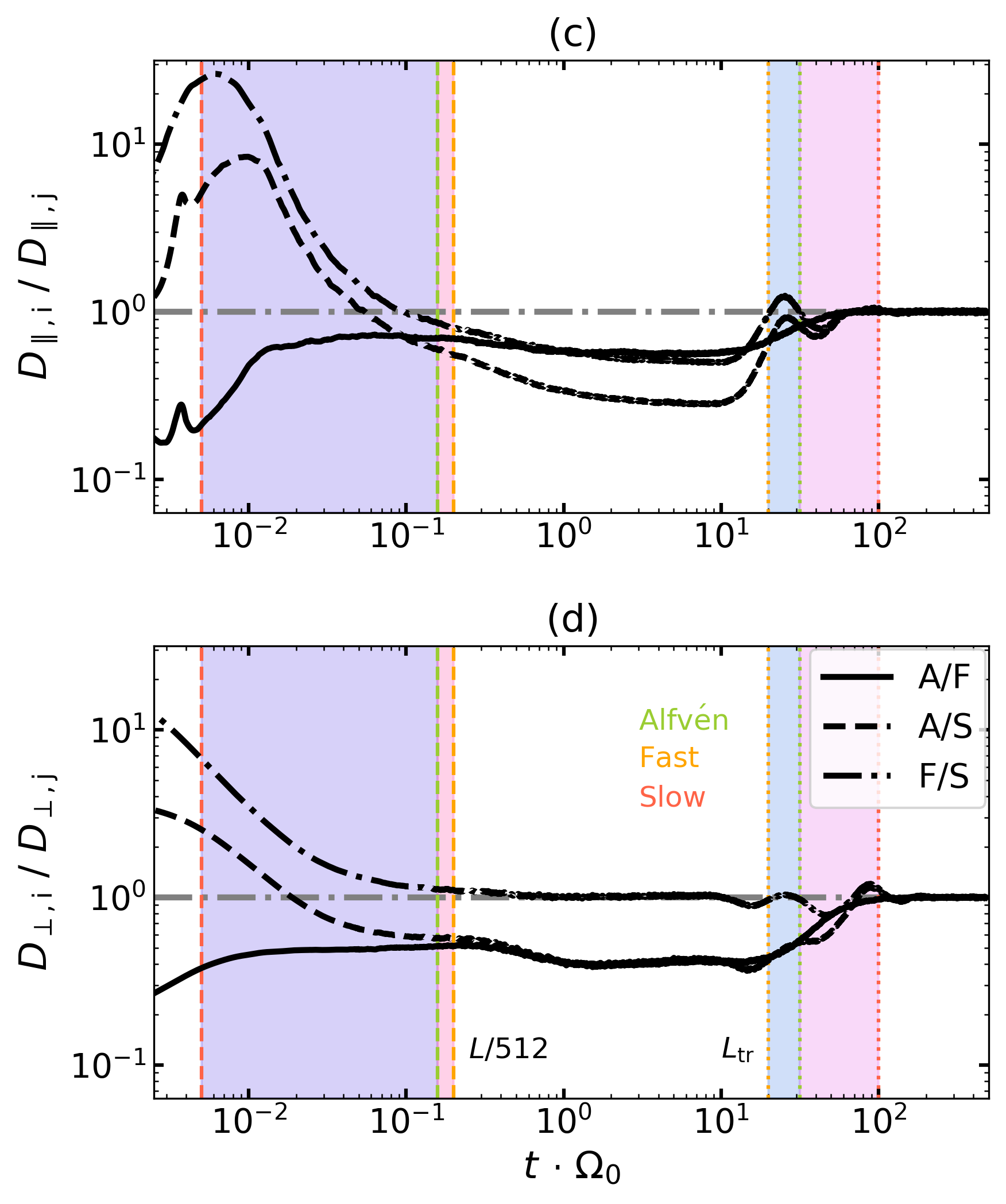}
\caption{The ratio of parallel (upper row) and perpendicular (lower) diffusion coefficients for three modes. The other descriptions are the same as those of Figure \ref{fig:08dpp2-afs}.
}
    \label{fig:09drr2-afs}
\end{figure*}

To explore the effect of the plasma modes on particle spatial diffusion, we present the ratio of parallel $D_{\parallel}$ (upper row) and perpendicular diffusion coefficient $D_{\perp}$ (lower row) between two modes in Figure \ref{fig:09drr2-afs}. As is shown in the left column (high $\beta$), the slow mode dominates the parallel diffusion $D_{\parallel}$ (panel a) during the whole simulation time, except for the time range after $\sim 10 /{\Omega_0}$ ($D_{\|, \rm A} \approx D_{\|,\rm F} \approx D_{\|,\rm S}$). From panel (b), we see that the dominance of $D_{\perp}$ among three modes can be roughly divided into two stages: (1) slow mode dominates perpendicular diffusion followed by $D_{\perp,\rm F} > D_{\perp,\rm A}$ before $\sim 10/{\Omega_0}$; (2) after $10 /{\Omega_0}$, magnetosonic modes dominate perpendicular diffusion with the relation of $D_{\perp,\rm F} \approx D_{\perp,\rm S} \approx D_{\perp, \rm A}$. It should be noted that the Alfv\'en mode plays a sub-dominant role almost in the whole evolution.

In the case of the low $\beta$ (right column), the ratio of parallel and perpendicular diffusion coefficients is almost similar. Before $\sim 0.1/{\Omega_0}$, both parallel $D_{\parallel}$ and perpendicular $D_{\perp}$ keep a dominant order, that is, fast mode is the largest followed by Alfv\'en mode, and slow mode is the smallest. From $\sim 0.1/{\Omega_0}$ to $\sim 10^{2}/{\Omega_0}$, the evolution of both $D_{\parallel}$ and $D_{\perp}$ in this situation is similar to the case of $D_{\parallel}$ in high $\beta$, where magnetosonic modes dominate the diffusion. After $\sim 10^{2}/{\Omega_0}$, their ratios are all close to 1, which means that the three modes play a comparable role. As a result, magnetosonic modes play a dominant role in the spatial diffusion of particles in the strong turbulence regime.

\begin{figure*}[t]
\centering
\includegraphics[width=0.48\textwidth,height=0.28\textheight]{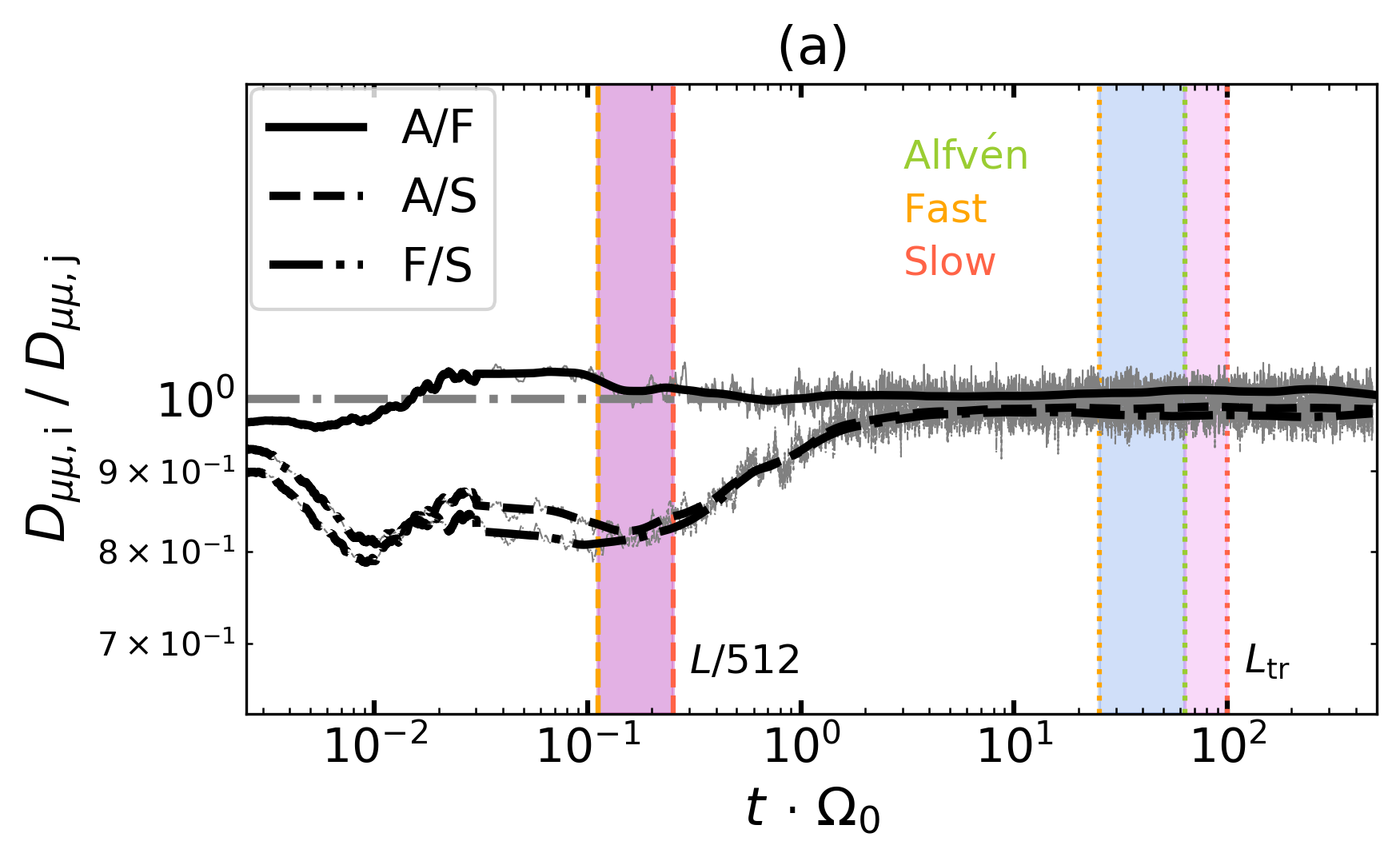}
\includegraphics[width=0.48\textwidth,height=0.28\textheight]{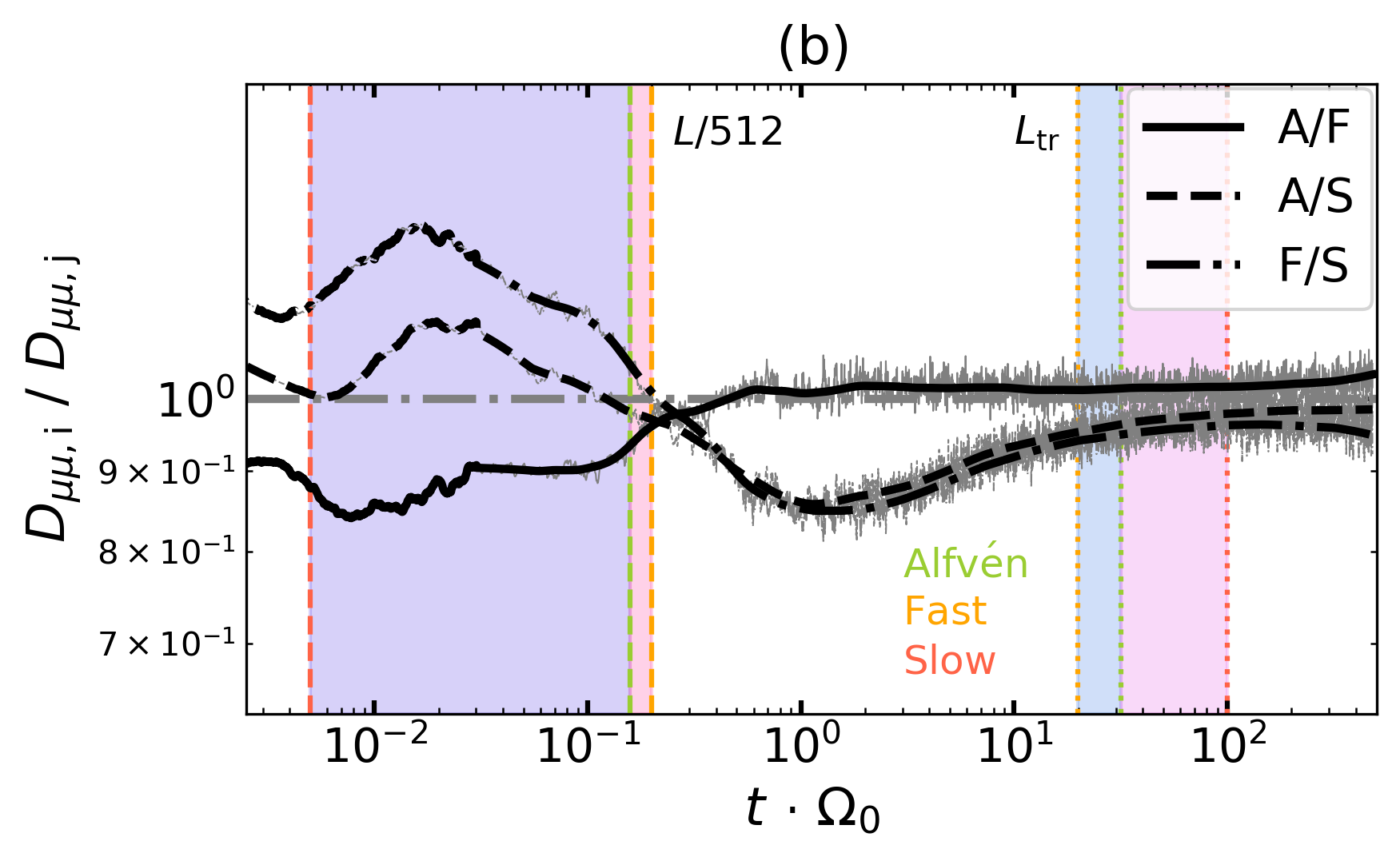}
\caption{The ratio of pitch-angle diffusion coefficients for three modes. The other descriptions are the same as those of Figure \ref{fig:08dpp2-afs}.
}
\label{fig:10dmumu2-afs}
\end{figure*}

$D_{\mu\mu}$ for three modes in Figure \ref{fig:10dmumu2-afs}. As is shown in panel (a), {i.e., the case of high $\beta$}, the dominant relation is $D_{\mu\mu, S} > D_{\mu\mu, F} \sim D_{\mu\mu, A}$ before $1/{\Omega_0}$ and then it turns to the stage that three modes approximately keep the similar scattering level. Differently, for the low $\beta$ regime (panel b), the fast mode is the largest, and Alfv\'en mode is secondary before $\sim 0.25/{\Omega_0}$, corresponding to $R_{\rm g}$ reaching the scale of $L/512$ for fast and Alfv\'en modes (see also Figure \ref{fig:07basicpara2}). After that, three ratios are approximately equal to 1 (after presenting a trough) consistent with the high $\beta$ scenario (see panel b). We would like to stress that when plotting, we constrain a narrow range of vertical coordinates to observe their differences. Therefore, using pitch-angle diffusion versus the evolution time, it is difficult to distinguish their differences from the scattering processes of the accelerated particles in the range of the box size.

\section{Discussion}\label{sec:discu}
In this work, we numerically explore the energization processes of CRs in compressible MHD turbulence regimes. We first focus on the influence of different turbulence properties on the CR energization and then on the interaction of CRs with Alfv\'en, slow, and fast modes. Specifically, we elaborate on the acceleration, diffusion, and scattering behavior of $10^4$ particles injected instantly by considering their evolution with the time and gyroradius.

Our current studies are built on a modern understanding of MHD turbulence theory, focusing on the properties of turbulence and the influence of plasma modes on particle acceleration. The results demonstrate that the diffusion coefficients of accelerated particles experience a universally exponential increase with the simulation time and gyroradius, similar to the phase of exponential growth seen in \cite{Demidem2020} for the relativistic MHD turbulence. Differently, \cite{Demidem2020} used the Monte Carlo simulation, one of the merits of which is that it is not affected by any choice of the distribution function. However, the limitation of this method is that its phenomenological description of acceleration processes cannot trace the microphysical details related to plasma physics.

When exploring the interplay of CRs and individual plasma modes, we find that the contribution to acceleration arising from the fast mode is the highest one in the high $\beta$ case, while the fast and slow modes dominate the acceleration in the low $\beta$ one, which is in good agreement with \cite{Cho2006} for the particle acceleration under strong and weak MHD turbulence. They theoretically derived the momentum diffusion coefficients of fast and slow modes and established that the acceleration efficiency of fast mode is more efficient than that of slow mode similar to the results of \cite{Yan2002} and \cite{Chandran2003}. The earlier studies have claimed that the scattering of charged particles by Alfv\'enic turbulence is negligible (\citealt{Yan2002, Yan2004, Lynn2014}), and magnetosonic modes are dominant processes for the transport and acceleration of CRs. However, we found that the effect of Alfv\'en mode cannot be ignored, playing a sub-dominant role. We conjecture that the non-negligible impact of the Alfv\'en mode on particle acceleration and diffusion may be from the compressible component of the Alfv\'en waves. In this regard, by projecting the local Alfven component of the turbulent variables into linear combinations of Alfv\'en and non-Alfv\'en components, \cite{Yuen2023} proposed an Alfv\'en leakage effect in the frame of the local magnetic field.

It is generally accepted that the mechanism by which turbulence causes particle acceleration is a stochastic acceleration process. Our current work does not describe what the acceleration mechanism is. As mentioned above, this paper focused on the influence of both turbulence properties and plasma modes on the transport of CRs. As well-known, the interaction of MHD turbulence would result in the gyroresonance and TTD of particles. For the former, its criterion is based on a comparison between wave frequency and particle Larmor frequency. At small scales, a particle can preserve its adiabatic invariant $\frac{mv_{\perp}^2}{2B_0} = const$, due to the Larmor radius being smaller than the variation scale of the magnetic field. However, particle motions would violate the adiabatic invariant conservation at large scales (\citealt{Chandran2000, Yan2003, Lazarian2021}). For the latter, it is essentially a Cerenkov-type interaction allowing particles to interact with large-scale turbulence. In the framework of the modern understanding of MHD turbulence, separating out the contributions of TTD and gyroresonance will be performed in future work.

When visualizing the time evolution of the gyroradius of particles (and also including the interaction of CRs with decomposed plasma modes), we use the weak-strong transition scale from theoretical predictions in LV99 and \cite{Lazarian2006} to explain our numerical findings. The theoretical prediction of weak-strong transition scales has been confirmed by simulations (\citealt{Verdini2012, Meyrand2016, Makwana2020}) and observations (\citealt{Sioulas2023, Zhao2023}). As an example, in the perspective of simulation, \cite{Meyrand2016} presented direct evidence of such a weak-strong transition, using a high-resolution three-dimensional direct numerical simulation of incompressible MHD turbulence. From the perspective of observation, \cite{Zhao2023} reported the first observational evidence for the Alfv\'enic weak-strong transition in MHD turbulence in Earth’s magnetosheath using data from the Cluster spacecraft.

In order to verify the reliability of our results, we also perform in this paper a comparison study using the numerical resolutions of $256^3$ and $792^3$. Our studies show that the difference in the numerical resolution does not affect the simulation results provided in the current work. Therefore, the resolution of 512 adopted in this paper is sufficient for our current goal. This work is only a first step towards understanding a more complex interplay between particle acceleration and plasma modes. 

\section{Summary}\label{sec:summa}
This paper studied the interaction of CRs with compressible MHD turbulence together with their acceleration processes based on the modern understanding of MHD turbulence theory. With different MHD turbulence regimes that may happen in a realistic astrophysical environment, we focused on particle acceleration, diffusion, and scattering processes using the test particle simulation.

\begin{enumerate}
\item We find that the gyroradius of particles exponentially increases with the simulation time by $R_{\rm g} \propto t^{\varphi}$. In the strong turbulence regime, the index $\varphi \in [4/3, 5/3]$ for sub-Alfv\'enic turbulence is steeper than $\varphi \in [1, 4/3]$ for super-Alfv\'enic one, while in the weak turbulence regime, $\varphi$ is approximately equal to $2/3$ for four turbulence cases explored.

\item In the strong turbulence range, the particle undergoes the superdiffusion in the momentum space, with the relationships of $D_{pp} \propto R_{\rm g}^{3/4}$ for sub-Alfv\'enic turbulence and $D_{pp} \propto R_{\rm g}^{2/5}$ for super-Alfv\'enic one. The momentum in the direction parallel to the local magnetic field dominates the diffusion process at large $R_{\rm g}$. In the weak turbulence regime, the momentum diffusion shows a plateau implying a stochastic acceleration process and meaning that the particle experiences the normal diffusion in the momentum space. 

\item With $R_{\rm g}$ in the range of box size, the parallel diffusion dominates the spatial diffusion of particles in the case of the sub-Alfv\'enic turbulence regime, while {in the weak turbulence regime} the perpendicular diffusion is slightly faster than the parallel one in the case of super-Alfv\'enic one. The pitch-angle diffusion decays along with the increasing gyroradius, with the diffusion rate in the weak turbulence slower than that in the strong turbulence.

\item As for the interaction of CRs with individual plasma modes, the property of particle acceleration, diffusion, and scattering is distinct from that of pre-decomposed MHD turbulence. The particle acceleration is dominated by the fast mode in the high $\beta$ case, while in the low $\beta$ case, it is dominated by the fast and slow modes. Moreover, magnetosonic (fast and slow) modes are also the main contributor to the momentum diffusion of CRs.

\item The spatial diffusion of particles is dominated by the slow mode for both the cases of high and low $\beta$ in the strong turbulence regime, while in the weak turbulence regime, three plasma modes play a comparable role. In particular, the spatial diffusion from the Alfv\'en mode cannot be ignored.

\end{enumerate}

\begin{acknowledgments}
We would like to thank the anonymous referee for constructive comments that have significantly improved our manuscript. We thank Alex Lazarian for reading throughout the manuscript, and Siyao Xu for helpful discussions on the particle acceleration and scattering in MHD turbulence. The authors thank the support from the National Natural Science Foundation of China (grant No. 11973035). J.F.Z. also thank the Hunan Natural Science Foundation for Distinguished Young Scholars (No. 2023JJ10039), and the China Scholarship Council for overseas research fund. 
\end{acknowledgments}

\bibliography{ms}{}
\bibliographystyle{aasjournal}

\end{document}